%% file: main_letter.tex
\documentclass[aps, prl, 10pt, twocolumn, superscriptaddress, noshowpacs, preprintnumbers, longbibliography, bibnotes,hyperref,floatfix]{revtex4-1}

\usepackage[colorlinks=true,breaklinks=true]{hyperref}
\hypersetup{allcolors=[rgb]{0.0 0.0 1.0},linkcolor=[rgb]{0.75 0.05 0.05}}
\usepackage[dvipsnames]{xcolor}
\usepackage{mathtools}
\usepackage{amsmath}
\usepackage{amsfonts}
\usepackage{amssymb}
\usepackage{bbold}
\usepackage{multirow}
\usepackage{bm}
\usepackage{hyperref}
\long\def\exclude#1{}
\usepackage{orcidlink}
\usepackage{slashed}
\usepackage[normalem]{ulem}
\usepackage{comment}

\usepackage{adjustbox}

\begin{document}

\title{Solar-System Abundances of $p$-Nuclides Probe\\ 
Collective Neutrino Oscillations in Supernovae}

\author{Alexander Friedland~\orcidlink{0000-0002-5047-4680}}
\email{alexfr@slac.stanford.edu}
\affiliation{SLAC National Accelerator Laboratory, Stanford University, Menlo Park, CA 94025}

\author{Derek J. Li~\orcidlink{0000-0001-6040-251X}\,}
\email{djpli@stanford.edu}
\affiliation{SLAC National Accelerator Laboratory, Stanford University, Menlo Park, CA 94025}
\affiliation{Leinweber Institute for Theoretical Physics, Stanford University, Stanford, CA 94305}

\author{Giuseppe Lucente~\orcidlink{0000-0002-3266-3154}}
\email{lucenteg@slac.stanford.edu}
\affiliation{SLAC National Accelerator Laboratory, Stanford University, Menlo Park, CA 94025}

\author{Payel Mukhopadhyay~\orcidlink{}}
\email{pm858@cam.ac.uk} 
\affiliation{SLAC National Accelerator Laboratory, Stanford University, Menlo Park, CA 94025}
\affiliation{Department of Physics, University of California Berkeley, Berkeley, CA 94720}
\affiliation{Department of Applied Mathematics and Theoretical Physics, University of Cambridge, United Kingdom
}

\author{Ian Padilla-Gay~\orcidlink{0000-0003-2472-3863}\,}
\email{ianpaga@berkeley.edu}
\affiliation{SLAC National Accelerator Laboratory, Stanford University, Menlo Park, CA 94025}
\affiliation{Department of Physics, University of California Berkeley, Berkeley, CA 94720}
\affiliation{Department of Physics, University of California San Diego, 
La Jolla, CA 92093}

\author{Amol V.\ Patwardhan~\orcidlink{0000-0002-2281-799X}}
\email{apatwardhan@reed.edu} 
\affiliation{SLAC National Accelerator Laboratory, Stanford University, Menlo Park, CA 94025}
\affiliation{%
School of Physics and Astronomy, University of Minnesota, Minneapolis, MN 55455
}
\affiliation{%
Department of Physics, New York Institute of Technology, New York, NY 10023
}
\affiliation{%
Department of Physics, Reed College, Portland, OR 97202
}

\preprint{SLAC-PUB-260622}
\preprint{N3AS-26-004}

\begin{abstract}

Direct evidence for collective neutrino oscillations in core-collapse supernovae remains elusive. We show that this quantum phenomenon leaves a footprint on the abundance pattern of proton-rich nuclides in the solar system. Modeling the $\nu p$-process using a $20\,M_\odot$ progenitor, we map 
out the dependence of the total yields on the starting radius of the oscillations, self-consistently coupling hydrodynamics and nucleosynthesis. The oscillations boost key $p$-nuclides 
($^{92,94}\text{Mo}$, $^{96,98}\text{Ru}$) and long-lived $^{92}\text{Nb}$ by 
up to two orders of magnitude, bringing their abundances into agreement with the observations. 
The best match is found when oscillations commence within $10\text{ km}$ 
of the proto-neutron star surface, indicating fast collective oscillations.

\end{abstract}

\maketitle

{\bf \emph{Introduction.}---}Just outside the protoneutron star (PNS) in a core-collapse supernova (CCSN), the dense gas of streaming neutrinos creates extreme conditions unattainable in terrestrial laboratories.  
Coherent $\nu$--$\nu$ scattering in this system couples the flavor states of neutrinos on different trajectories, driving \emph{collective flavor oscillations}~\cite{Pantaleone:1992eq, Sigl:1992fn, McKellar:1992ja, Samuel:1993uw, Qian:1994wh,  Duan:2010, Mirizzi:2015eza, Chakraborty:2016yeg, Tamborra:2020cul, Capozzi:2022slf, Richers:2022zug, Volpe:2023met,Johns:2025mlm}. Analytical and numerical studies of this many-body quantum phenomenon 
have uncovered very rich physics, including \emph{fast collective oscillations} (FCO)~\cite{Sawyer:2005jk, Sawyer:2008zs, Sawyer:2015dsa, Dasgupta:2016dbv, Izaguirre:2016gsx, Chakraborty:2015tfa, Chakraborty:2016lct, Nagakura:2019sig,Morinaga:2019wsv,Abbar:2019zoq,Glas:2019ijo,Nagakura:2021hyb,Padilla-Gay:2021haz,Zaizen:2022cik, Nagakura:2022xwe,Xiong:2024tac,Fiorillo:2024pns, Fiorillo:2025ank} driven purely by $\nu$--$\nu$ coherent scattering and \emph{slow collective oscillations} (SCO)~\cite{Kostelecky:1994dt, Samuel:1995ri, Pastor:2002we, Duan:2005cp, Duan:2006an, Duan:2006jv, Hannestad:2006nj, Dasgupta:2007ws, Duan:2008za, Friedland:2010sc, Duan:2010bf, Padilla-Gay:2025tko, Fiorillo:2024bzm, Fiorillo:2024uki} driven by the interplay between the $\nu$--$\nu$ coherent scattering and neutrino-mass-induced transformations. Yet, a complete theoretical description of this phenomenon in realistic CCSN conditions remains elusive, due to the extreme complexity of the problem rooted in the nonlinearity of oscillations, the presence of disparate length- and time-scales~\cite{Bell:2003mg, Friedland:2003eh, Friedland:2006ke}, and essential multi-dimensional dynamics~\cite{Mezzacappa:2020oyq, Johns:2025mlm}. It is then crucial to identify any observables that carry imprints of FCO or SCO, especially those that are sensitive to how far from the PNS surface the oscillations develop.   In this \emph{Letter}, we argue that the desired probe is provided by the pattern of certain proton-rich isotopes observed in the Solar System. Moreover, we show that the oscillations may be \emph{necessary} to explain the observations.

It has long been known that if collective oscillations were to occur in the hot bubble surrounding the PNS, they could impact nucleosynthesis there~\cite{Qian:1993dg,Pantaleone:1994ns,Pastor:2002we}. These early papers considered the $r$-process (rapid neutron capture)  and invoked sterile neutrinos. It has since been shown, however, that collective oscillations can occur in the hot bubble even with only active neutrinos~\cite{Duan:2005cp}. Combining representative oscillation scenarios with nucleosynthesis calculations then showed that the $r$-process yields can depend sensitively on where flavor mixing starts~\cite{Duan:2010af}. 
This makes nucleosynthesis a promising complementary probe to neutrino signal detected on Earth, as the latter only reflects the final state of the neutrino evolution. 

Our understanding of the type of nucleosynthesis that can occur in the hot bubble has also advanced. The $\nu_e$ and $\bar\nu_e$ spectra predicted by modern CCSN simulations tend to yield \emph{proton-rich} conditions in the neutrino-driven outflow from the PNS surface that forms the hot bubble, 
especially once the effect of weak magnetism and recoil~\cite{Horowitz:1999fe, Horowitz:1999wy, Horowitz:2001xf} is accounted for~\footnote{This subdominant but important effect boosts the $\nu_e$ cross sections on nucleons compared to those of $\bar\nu_e$}.
This is, in fact, a welcome development. While the $r$-process may efficiently proceed in neutron star binary mergers~\cite{Arnould:2007gh, Kasen:2017sxr, Thielemann:2026EPJA}, one also needs to identify a site where $p$-nuclides are made. The origin of these isotopes, especially $^{92,94}$Mo and $^{96,98}$Ru, which are rather abundant in the Solar System, has been a long-standing mystery~\cite{ArnouldGoriely2003, Rauscher:2013}. The hot bubble in a CCSN turns out to be a promising site, supplying not only the proton-rich conditions, but also a large flux of streaming electron antineutrinos. The latter helps solve~\cite{Frohlich:2005ys,Pruet:2005qd,Wanajo:2006rp} the famous ``waiting points problem'' of the classic $rp$-process~\cite{1999ApJ...524.1014S}. This neutrino-assisted mechanism is called the $\nu p$-process~\cite{Frohlich:2005ys}.

While there are several recent notable investigations of the impact of FCO on the CCSN explosion mechanism~\cite{Ehring:2023abs, Ehring:2023lcd, Nagakura:2023mhr, Wang:2025nii, Wang:2025ihh, Mori:2025cke, Gogilashvili:2026kef, Gogilashvili:2026epg, Akaho:2026kff}, there is only a single study on how the $\nu p$-process is affected by FCO~\cite{Xiong:2020ntn}. There is, likewise, a limited number of studies exploring the effects of SCO on the $\nu p$-process~\cite{Martinez-Pinedo:2011yhi,Sasaki:2017jry,Balantekin:2023ayx}. 
While these papers contain important findings, they do not treat the impact of the flavor transformations on the hydrodynamics self-consistently, which is required to tackle the problem. 

\begin{figure}[t!]
    \centering
    \includegraphics[width=\columnwidth]{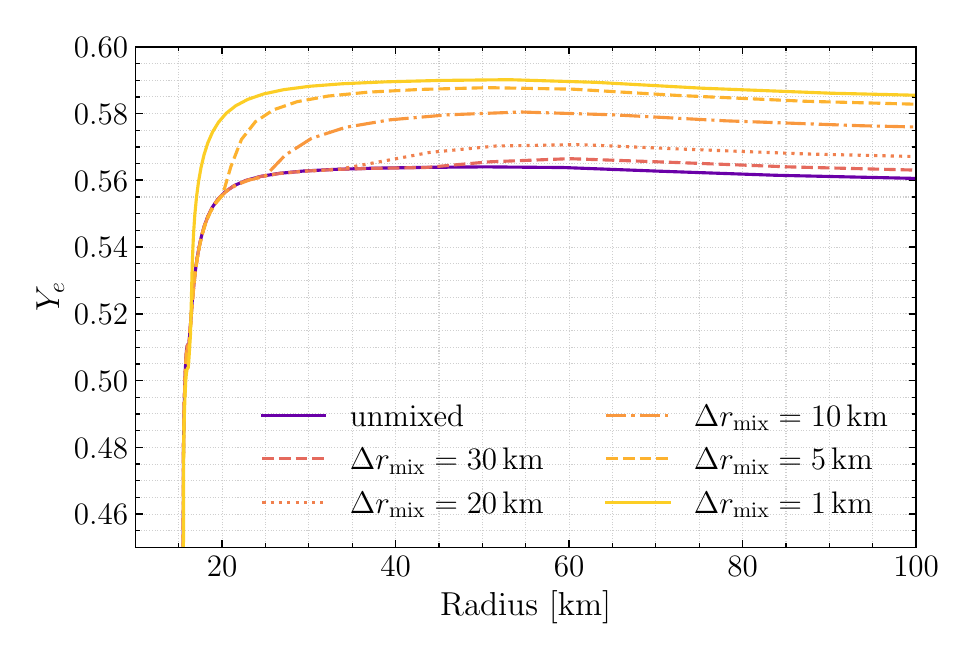}
    \caption{
    Electron fraction $Y_e$ as a function of radius for our benchmark $20\,M_\odot$ model at $t=3.5$\,s, shown for the outflow without neutrino mixing and with mixing occurring at various distances $\Delta r_{\rm mix}$ from the PNS surface.}
    \label{fig:Ye}
\end{figure}


{\bf \emph{Analysis requirements.}---}Flavor oscillations change the spectra of $\nu_e$ and $\bar\nu_e$, by mixing them with the original spectra of $\nu_x$ and $\bar\nu_x$, with $x=\mu,\tau$. 
This impacts the $\nu p$-process in three ways. First, it changes the rates of neutron-producing reactions $\bar\nu_e+p\rightarrow e^+ + n$ during the proton-capture phase of the $\nu p$-process. This is a \textit{far} effect of the oscillations, as it occurs hundreds of km from the PNS surface~\cite{Xiong:2020ntn,Martinez-Pinedo:2011yhi,Sasaki:2017jry,Balantekin:2023ayx}. Second, oscillations modify the electron fraction $Y_e$ in the outflow, set by the reactions $\bar\nu_e+p\rightarrow e^+ + n$ and $\nu_e+n\rightarrow e^- + p$. Since the unoscillated energy hierarchy is typically larger for $\nu_e$–$\nu_x$ than $\bar\nu_e$–$\bar\nu_x$, mixing hardens $\nu_e$ more than $\bar\nu_e$, enhancing $\nu_e+n\to p+e^-$ relative to $\bar\nu_e+p\to n+e^+$ and increasing $Y_e$~\cite{Xiong:2020ntn}. This is only relevant if the oscillations occur within {10-20~km} from the PNS surface, where $Y_e$ is set (see Fig.~\ref{fig:Ye}). It is thus considered a \textit{near} effect. Third, as another \textit{near} effect, flavor conversion increases the heating rate $\dot{Q}$~\cite{Xiong:2020ntn}. 

The last effect energizes the outflow and deserves special consideration. Two key physical factors enter here: (i) Effective $\nu p$-process requires a certain matching between the processes that occur at different stages of the outflow~\cite{Wanajo:2010mc,Friedland:2023kqp}. The number of seeds produced during the stage of quasi-statistical equilibrium ($3\,{\rm GK} \lesssim T \lesssim 6\,{\rm GK}$) has to be in a certain ratio with the number of neutrons produced in the $\bar{\nu}_e(p,\,n)e^+$ reactions in the $p$-nuclide formation stage ($1.5\,{\rm GK} \lesssim T \lesssim 3\,{\rm GK}$). Refs.~\cite{Friedland:2023kqp, Friedland:2025lge} established the hydrodynamic conditions for optimal yields: a heavy PNS mass and a subsonic outflow persisting for several seconds. (ii) As argued in~\cite{Friedland:2020ecy}, the outflow in a CCSN, under realistic conditions, can be either subsonic or supersonic, depending on a combination of relevant parameters. Because of this near-critical nature of the outflow, the extra heating induced by neutrino oscillations might cause supersonic transitions, depressing the yields and counteracting the effect of boosted $Y_e$. To assess the effect of flavor mixing on the yields then \emph{requires} us to recompute the hydrodynamics of the outflow for each oscillation scenario.

{\bf \emph{Methods.}---}We use as benchmark a spherically symmetric long-term supernova simulation of a $20\,M_\odot$ progenitor with a $1.93\,M_\odot$ PNS from the {\tt GARCHING} CCSN Archive~\cite{SNarchive,Fiorillo:2023frv,Lucente:2024ngp}. This state-of-the-art simulation meets the hydrodynamic requirements for efficient $\nu p$, as outlined above, yet was not tailored for the $\nu p$-process by its authors. We take as input the PNS mass $M$, time-dependent PNS radius, (unoscillated) neutrino luminosities and spectra, and the progenitor profile~\cite{Sukhbold:2015}. We model the luminosity evolution with the fit from Ref.~\cite{Lucente:2024ngp}, and we fix spectral moments at their values at post-bounce time $t=3.5$\,s in the SN simulation  (when the $\nu p$-process peaks), assuming pinched Fermi-Dirac distributions~\cite{1989A&A...224...49J,1989A&AS...78..375J,Keil:2002in}.

As mentioned above, we are compelled to recompute the hydrodynamics for each oscillation scenario. We do so with the framework of Ref.~\cite{Friedland:2025lge}, which solves general-relativistic steady-state hydrodynamics in spherical symmetry, with an equation of state including the baryon gas and variable radiation degrees of freedom. Neutrino interactions with both nucleons and electrons are included, and \textit{time-dependent} boundary conditions at the PNS surface and the hot-bubble's edge (with proper merging into the colder outflow) are imposed, as discussed in Supplemental Material (SM) A, B. 
We construct the time evolution through a sequence of snapshots, representing particles launched 0.1\,s apart.

Following existing literature~\cite{Xiong:2020ntn}, we model flavor conversion as complete instantaneous equilibration at a distance $\Delta r_{\rm mix}$ from the PNS surface. Beyond this radius, we replace the original spectra with the equilibrated ones, combining all flavors equally across the energy modes: $f'_{\nu_e(\bar\nu_e)}=(f_{\nu_e(\bar\nu_e)}+2f_{\nu_x(\bar\nu_x)})/3$, thereby modifying neutrino energy moments and luminosities (see SM~C).\,\footnote{This prescription should be viewed as a representative, approximate limiting case, since flavor equilibration is constrained by lepton-flavor number conservation \cite{Dasgupta:2017oko,Abbar:2018beu}. However, the resulting asymptotic value of $Y_e$ is likely to vary within a narrow range around $\sim$ 0.6. A further quantitative investigation will follow in future work~\cite{Friedland2026}.} This modifies both neutrino heating (thus hydrodynamics) and nucleosynthesis. We vary $\Delta r_{\rm mix}$ as a free parameter to quantify the $\nu p$-process's sensitivity to conversion location.

\begin{figure*}[t!]
    \centering
    \includegraphics[width=1.\columnwidth]{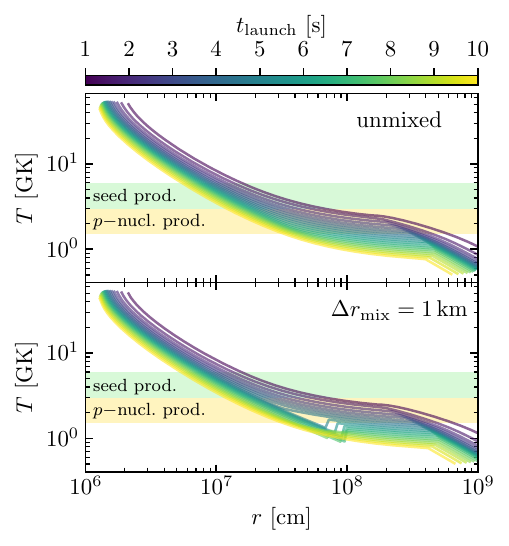}
    \includegraphics[width=1.\columnwidth]{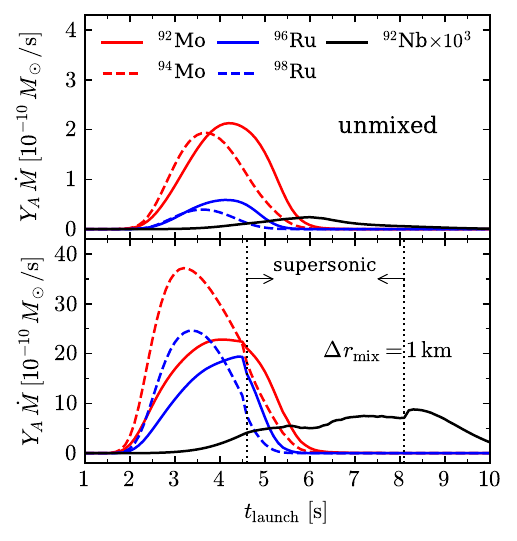}
    \caption{\emph{Left:} Radial temperature profiles for outflows launched at $t_{\rm launch}\in[1,\,10]\,{\rm s}$, depicted here in steps of $0.5\,{\rm s}$. The green and yellow bands correspond to the seed-formation and $p$-nuclide formation stages of the $\nu p$-process, respectively~\cite{Wanajo:2010mc,Friedland:2025lge}. \emph{Right:} Time evolution of the production rates of Mo (red) and Ru isotopes (blue), and $^{92}{\rm Nb}$ (black), with vertical dotted lines delimiting the time window where the outflows contain a supersonic transition. Top (bottom) panels show the unmixed case (flavor equilibration occurring $1$ km above the PNS surface).} 
    \label{fig:time_evolution}
\end{figure*}

Nucleosynthesis is computed by post-processing tracer trajectories with the reaction network {\tt SkyNet}~\cite{Lippuner:2017tyn, skynet}, incorporating modifications from Ref.~\cite{Friedland:2025lge}  as detailed in SM~A (including medium-enhanced triple-$\alpha$~\cite{Beard:2017jpg, Jin:2020}). We further modify the code to include distance-dependent flavor mixing (SM~C). Each trajectory is launched at $t_{\rm launch}\in[1,\,10]\,$s after the core bounce and evolves for $\sim30$\,years in {\tt SkyNet}, with conditions evaluated at $t=t_{\rm launch}+t_{\rm {\tt sky}}$, where $t_{\rm {\tt sky}}$ is {\tt SkyNet}'s evolution time (starting from zero at launch). Beyond the benchmark model, we perturb the PNS and progenitor's properties to test the robustness of our qualitative findings across different regimes.

{\bf \emph{Oscillation-driven physical effects.}---}The impact of neutrino oscillations on the $\nu p$-process depends on {\it how far} from the PNS flavor mixing takes place. If conversions occur at $\Delta r_{\rm mix} \gtrsim 40$\,km (``far conversions''), only the rate of neutron production during the proton-capture phase is appreciably changed, while $Y_e$ and $\dot{Q}$ are not affected. For the model we use in this paper, the $\bar{\nu}_e(p,n)e^+$ rates are enhanced by $\sim 20\%$.

Conversions at smaller radii (``near conversions'') induce all three effects simultaneously: $\sim 20\%$ enhancement of $\bar{\nu}_e(p,n)e^+$ rates, up to $\sim 30\%$ increase in $\dot{Q}$ (and also the mass outflow rate $\dot{M}$), and boosted $Y_e$. The impacts on $Y_e$ and $\dot{Q}$ exhibit similar dependence on $\Delta r_{\rm mix}$ (see SM~C).

Figure~\ref{fig:Ye} shows $Y_e(r)$ at a representative post-bounce time for several values of the flavor equilibration radius. Notice that the important weak-magnetism and recoil corrections~\cite{Horowitz:1999fe, Horowitz:1999wy, Horowitz:2001xf} are included here and in what follows. With the original neutrino spectra, $Y_e$ approaches an equilibrium value of $\approx 0.56$ (solid violet). Neutrino mixing increases $Y_e$, rendering the outflow more proton-rich, consistent with Ref.~\cite{Xiong:2020ntn}. This enhancement is stronger the closer mixing occurs to the PNS. For $\Delta r_{\rm mix}=1\,$km, $Y_e$ increases up to $\approx 0.59$ (solid yellow), while there is negligible impact for $\Delta r_{\rm mix}\gtrsim 30$\,km.  

Notice that earlier studies found $Y_e \approx 0.6$ to be conducive to  successful $\nu p$-process, e.g.~\cite{Wanajo:2010mc,Friedland:2023kqp,Friedland:2025lge}. That flavor mixing brings $Y_e$ close to this desired value is, in fact, not an accident of the specific simulation we are considering and rather insensitive to the details of flavor conversion. As can be readily shown (see SM~A), in the limit when the $\nu_e$ and $\bar{\nu}_e$ spectra are identical, $Y_e$ reaches $\approx 0.63$. Since the original $\nu_x$ and $\bar{\nu}_x$ spectra are close, any flavor mixing dynamics that transfers substantial $\nu_x$ flux into $\nu_e$ boosts $Y_e$ to near the desired value $\sim 0.6$. 
For instance, were the outcome of near conversion to be
spectral swaps of the kind studied in
Refs.~\cite{Duan:2006jv,Duan:2007bt,Raffelt:2007xt,Dasgupta:2009mg,
Friedland:2010sc,Dasgupta:2010cd,Duan:2010bf}, the resulting $Y_e$ shift
would be comparable.

{\bf \emph{Results of near flavor conversion.}---}Oscillation-induced changes to $Y_e$ (Fig.~\ref{fig:Ye}) and $\dot Q$ (explicitly shown in SM~C) are largest as flavor conversion occurs close to the PNS. Thus, we begin by comparing two extreme cases: no mixing and flavor equilibration occurring $1$ km above the PNS surface ($\Delta r_{\rm mix}=1$\,km).

\begin{figure*}[t!]
    \centering
    \includegraphics[width=0.99\columnwidth]{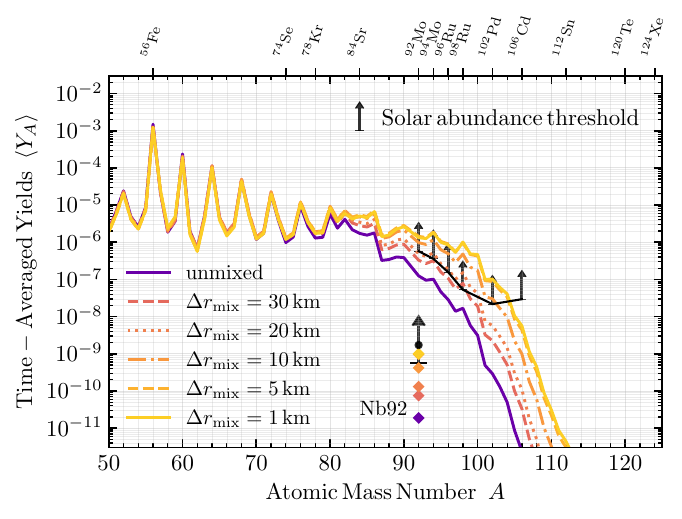}
    \includegraphics[width=0.99\columnwidth]{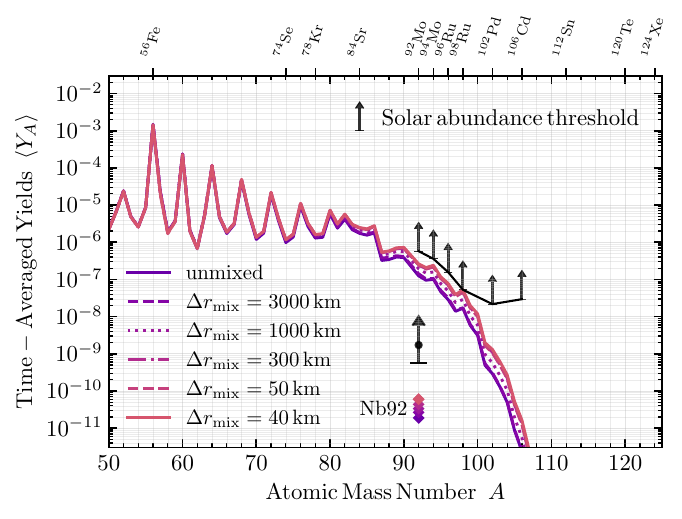}
    \caption{Time-averaged yields $\langle Y_A\rangle$ for our benchmark $20\,M_\odot$ model 
     under flavor conversion occurring at various radii from the PNS.
     Curves show $\langle Y_A\rangle$ versus mass number $A$; diamonds mark the corresponding $^{92}$Nb abundance.
     \emph{Left:} ``Near'' conversion, i.e., mixing within $30$\,km from the PNS surface, which modifies $Y_e$ and $\dot Q$ and also enhances neutron production rates.
    \emph{Right:} ``Far'' conversion, i.e., mixing at larger radii, affecting only neutron production during the $p$-nuclide formation stage.
    Black arrows indicate the minimum abundances needed to reproduce the solar values (see main text and SM).}
    \label{fig:Integrated_yields}
\end{figure*}

The left panels of Fig.~\ref{fig:time_evolution} show the evolution of the outflows launched at different $t_{\rm launch}$ for our benchmark $20\,M_\odot$ model. Without mixing, the outflow remains subsonic throughout the cooling phase (upper left). For $\Delta r_{\rm mix}=1$\,km (lower left), the additional heating induces a transonic transition at $t_{\rm launch}\gtrsim 4.5$\,s, and the termination shocks, visible as discontinuities in the radial temperature profiles around $r\simeq 10^8$\,cm, persist until $t_{\rm launch}\approx 8$\,s, as the luminosities decline. Such emergence of transonic outflows from a modest increment of heating illustrates the near-critical nature of CCSN hydrodynamics~\cite{Friedland:2020ecy}, and, practically, the necessity of self-consistently solving the hydrodynamics for small $\Delta r_{\rm mix}$.

The right panels of Fig.~\ref{fig:time_evolution} show the corresponding production rates as a function of time of several $p$-nuclides and the long-lived radionuclide $^{92}$Nb, a key diagnostic of the $\nu p$-process's efficiency~\cite{Rauscher:2013,Friedland:2023kqp,Friedland:2025lge}.\footnote{Studies prior to Refs.~\cite{Friedland:2023kqp,Friedland:2025lge} used insufficient $^{92}$Nb production to argue against $\nu p$-process viability~\cite{Rauscher:2013,Lugaro:2016zuf,Haba:2021PNAS}.} Without mixing (upper right), $p$-nuclides are efficiently produced in an optimal window of $\sim 3\text{--}4$\,s in the early cooling phase, while $^{92}{\rm Nb}$ is produced later. With $\Delta r_{\rm mix}=1$\,km (lower right), the synthesis of Mo and Ru isotopes is enhanced by factors of $\sim10$--$20$ and $\sim 30$--$60$, respectively, reflecting higher $Y_e$ and (anti)neutrino capture rates. Importantly, the supersonic transition sets in only after the production rates have peaked (vertical dotted line at $t_{\rm launch}\simeq4.5$\,s). While the onset of mildly supersonic conditions noticeably accelerates the post-peak decline of Mo and Ru nucleosynthesis, the production rates remain higher than the original (unmixed) case at each $t_{\rm launch}$. Oscillations increase $^{92}\mathrm{Nb}$ production by as much as two orders of magnitude. In contrast to $p$-nuclides, $^{92}$Nb has a significantly broadened production window and distinct temporal behavior.

{\bf \emph{Yields as a function of mixing radius.}---}To compare with observations, we next compute the total yields in a given explosion by integrating over time. We present our results as time-averaged yields normalized by the total mass ejected
\begin{equation}
\langle Y_{A,Z}\rangle = \frac{\int dt\, Y_{A,Z}(t)\dot{M}(t)}{\int dt\, \dot{M}(t)}\,.
\end{equation}
With this, we systematically investigate the dependence of $\langle Y_{A,Z}\rangle$ on the mixing location, the first time such a study is undertaken in the literature. Figure~\ref{fig:Integrated_yields} shows $\langle Y_A \rangle =\sum_{Z} \langle Y_{A,Z} \rangle$ versus mass number $A$ for near (left) and far (right) conversions for different values of $\Delta r_{\rm mix}$. For the $p$-nuclides labeled on the upper axis, $\langle Y_A \rangle$ corresponds to $\langle Y_{A,Z}\rangle$ of the indicated isotope, the only relevant species at that $A$ produced by the $\nu p$-process. 

From the left panel of Fig.~\ref{fig:Integrated_yields}, we see that the largest enhancement is obtained when $\Delta r_{\rm mix}<10\,$km. Here, for $90\le A\le 100$, flavor equilibration enhances the yields up to $\sim 2$ orders of magnitude relative to the unmixed case (solid violet) for both $p$-nuclides (curves) and $^{92}$Nb (diamonds), with factors of $\sim 10$ for $^{92}$Mo and $\sim 60$ for $^{98}$Ru. This reflects the enhancement in the production rates illustrated in Fig.~\ref{fig:time_evolution}. Counterintuitively, despite larger $Y_e$ values at smaller $\Delta r_{\rm mix}$ (Fig.~\ref{fig:Ye}), the cases with $\Delta r_{\rm mix}=1,\,5,\,10$\,km yield similar $\langle Y_A\rangle$. This is due to the competing effect of the increase in heating: mixing closer to the PNS triggers an earlier transonic transition that shortens the optimal production window, partially offsetting the $Y_e$-driven gain. 

 For $10\,{\rm km} \lesssim \Delta r_{\rm mix}\lesssim 40\,{\rm km}$ the $\nu p$-yields are strongly sensitive to the value of $\Delta r_{\rm mix}$. Here, no supersonic transition takes place, and the effect of $Y_e$ is not counterbalanced by hydrodynamics. As expected, the yields fall monotonically with increasing $\Delta r_{\rm mix}$, approaching a factor of $2\text{--}3$ enhancement relative to the unmixed case for $\Delta r_{\rm mix} =40$\,km (see SM~D).

For $40\,{\rm km} \lesssim \Delta r_{\rm mix} \lesssim 300$\,km (right panel of Fig.~\ref{fig:Integrated_yields}) the yields are nearly independent of $\Delta r_{\rm mix}$. In this regime, $Y_e$ and $\dot Q$ are effectively not modified, leaving the far effect as the only relevant process, which operates at the $p$-nuclide production stage, starting roughly at 300 km (see Fig.~\ref{fig:time_evolution}, top-left panel). At larger $\Delta r_\text{mix}$, enhancement diminishes, becoming negligible for $\Delta r_{\rm mix}\gtrsim 3000$\,km where $p$-nuclide production ends.

{\bf \emph{Comparison with solar abundances.}---}Black arrows in Fig.~\ref{fig:Integrated_yields} mark solar-abundance thresholds from meteoritic data~\cite{Lodders:2003} for $^{92}$Mo--$^{106}$Cd. These are computed by requiring that $\nu p$-process mass fractions, when diluted in the total ejecta, exceed solar values by one order of magnitude~\cite{Wanajo:2010mc,Friedland:2025lge} (see SM~D). The unmixed case falls short by approximately one order of magnitude. Far conversions, increasing yields by only factors of a few, remain insufficient to reach solar thresholds. In contrast, near conversions enable agreement with solar abundances~\cite{Wanajo:2010mc,Friedland:2025lge}: our model successfully meets solar requirements of the key Mo and Ru isotopes for $\Delta r_{\rm mix}\lesssim 20$--$30$\,km, while for $\Delta r_{\rm mix}\lesssim 10$\,km, $^{92}$Nb, as well as $^{102}$Pd, become viable.

{\bf \emph{Discussion.---}}Near flavor conversion makes our $20\,M_\odot$ model an efficient $\nu p$-process site. Modest perturbations of the benchmark parameters preserve our qualitative results: when conversion occurs sufficiently close to the PNS surface, yields reach the solar thresholds, provided that the PNS and progenitor remain sufficiently massive. This is demonstrated for a representative case in SM~E, obtained by reducing the progenitor mass to $18.6\,M_\odot$ and modestly increasing the front shock velocity to 7000 km/s. In this model, lower confining pressure triggers an earlier transonic transition, clipping the optimal window even without neutrino mixing. Flavor conversion leads to similar enhancement factors of the yields: for $\Delta r_{\rm mix}=1$\,km, the transonic transition occurs $\sim 1$\,s earlier, yet an early subsonic interval persists wherein boosted $Y_e$ and $\dot{M}$ enhance production. Though reduced relative to the benchmark, yields of $^{92,94}$Mo, $^{96,98}$Ru, and $^{92}\mathrm{Nb}$ still exceed solar thresholds for $\Delta r_{\rm mix}\lesssim 10$\,km.

More generally, flavor mixing enhances yields in all cases we explored, even when (mild) transonic transitions already occur near the beginning of the cooling phase.
However, the $\nu p$-process is inefficient in strongly supersonic conditions~\cite{Friedland:2023kqp,Friedland:2025lge}, rendering mixing-induced enhancements insufficient to reach solar thresholds. An exploration of different models is left for future work~\cite{Friedland2026}.

Our systematic study establishes that \textit{near} conversion---triggered by FCO (see e.g.,
Refs.~\cite{Tamborra:2020cul,Capozzi:2022slf,Volpe:2023met} for recent reviews) within $\sim 30$ km from the PNS surface~\cite{Nagakura:2019sig,Morinaga:2019wsv,Abbar:2019zoq,Glas:2019ijo,Nagakura:2021hyb}---increases yields by more than one order of magnitude, allowing agreement with solar observations for sufficiently massive progenitors and PNS. By contrast, \textit{far} conversion, associated with SCO at $40\text{--}500$ km~\cite{Duan:2006jv,Duan:2006an,Duan:2005cp,Hannestad:2006nj}, enhances yields by only factors of a few. Mixing at larger distances, i.e. $\Delta r_{\rm mix}\sim \mathcal{O}(10^3)$\,km, has negligible impact\,\footnote{Oscillations at these distances might arise from the Mikheyev-Smirnov-Wolfenstein (MSW) effect, a manifestation of the coherent forward scattering of neutrinos with the electrons in the medium~\cite{Wolfenstein:1977, Mikheev:1986}. This becomes only relevant when the outflow reaches densities of about $10^3 \ {\rm g/cm^3}$ at distances of thousands of kilometers from the PNS. These densities are reached, however, at late times $t \gtrsim 6 $~s~\cite{Friedland:2020ecy} i.e. after the optimal window of the $\nu p$-process.}. These results suggest that FCO may play a crucial role in establishing the observed abundances of light $p$-nuclides in the solar system.

The relative importance of far and near conversions depends on the closeness of unoscillated $\bar{\nu}_e$ and $\bar{\nu}_x$ spectra in both energy and luminosity during cooling. Similar spectra are commonly observed in state-of-the-art simulations ~\cite{Fiorillo:2023frv,Janka:2025tvf,Fischer:2011cy,Nakazato:2012qf,Fischer:2023ebq,Choi:2025igp}, while large differences typically occur in simulations neglecting inelastic neutrino-nucleon scattering corrections~\cite{Sasaki:2017jry} (see Fig.~14 in Ref.~\cite{2012ApJ...756...84M}). The {\tt GARCHING} group finds an average energy difference  $\Delta \langle E_{\bar{\nu}} \rangle \equiv \langle E_{\bar{\nu}_x}\rangle - \langle E_{\bar{\nu}_e}\rangle \lesssim 0.5$\,MeV~\cite{Fiorillo:2023frv,Janka:2025tvf}, slightly smaller than the one observed by other groups ($\Delta \langle E_{\bar{\nu}} \rangle \lesssim 1$--$2$\,MeV)~\cite{Fischer:2011cy,Nakazato:2012qf,Fischer:2023ebq,Choi:2025igp}. Our far-conversion results agree qualitatively with Ref.~\cite{Martinez-Pinedo:2011yhi}, which assumes $\Delta \langle E_{\bar{\nu}}\rangle \approx 0.9$\,MeV based on simulations from Ref.~\cite{Buras:2005rp}. By contrast, adopting greater $\bar\nu_e$--$\bar\nu_x$ spectral differences ($\Delta\langle E_{\bar\nu}\rangle \gtrsim 5$\,MeV), Refs.~\cite{Sasaki:2017jry, Balantekin:2023ayx} found substantially larger enhancements in $\bar\nu_e$-capture rates and $\nu p$-yields. Thus, the impact of far conversion is sensitive to the details of neutrino transport, which motivates careful future investigations, given its importance to the estimation of $\nu p$-yields in the presence of oscillations.

{\bf \emph{Conclusions.}---}
We saw that flavor oscillations close to the PNS can transform the nature of the outflow and with it the efficiency of the $\nu p$-process. This requires the recomputation of the hydrodynamics for each oscillation scenario, while also including the oscillation effects on the electron fraction and the neutron production rate. Such a self-consistent treatment of the hydrodynamics and nucleosynthesis, as well as the systematic investigation of the dependence of the yields on the starting radius of the oscillations, are presented here for the first time.

Our calculations confirm that the yields of the $\nu p$-process are sensitive to the starting radius of collective oscillations. We find that the impact of the oscillations is greatest when they occur within $\sim10$ km of the PNS surface, in which case the yields of the $p$-nuclides in the crucial $A = 90\text{--}100$ mass range are boosted by up to 20-60 times. In comparison, when the neutrino mixing radius is $\gtrsim40$ km, the enhancement is an order of magnitude smaller and, moreover, the sensitivity to the exact location of the oscillations is reduced. 

Crucially, when the oscillations occur close to the PNS surface, \emph{and only then}, the predicted abundances line up with the solar-system observations. This applies to all $p$-isotopes up to $^{102}$Pd, as well as to $^{92}$Nb, which also serves as a probe of the efficiency of the $\nu p$-process. Thus, collective oscillations may be responsible for the observed pattern of these isotopes. Notably, the agreement is reached entirely within the physical conditions of a modern CCSN simulation, without tuning any aspects of explosion, such as the electron fraction, neutrino fluxes, entropy per baryon, or the outflow profile, the first such example in the literature.

Given today's best available information on the dynamics of collective oscillation and on the spectra of different neutrino flavors, our results might be taken as an indication that the relevant mechanism is fast collective. Further investigations incorporating collective oscillation dynamics self-consistently into CCSN simulations are needed to investigate the robustness of this finding.

\acknowledgements

\textbf{\textit{Acknowledgments}}---The work of 
all authors at SLAC was supported by the U.S. Department of Energy under contract number DE-AC02-76SF00515. AVP acknowledges partial support from the U.S. Department of Energy under contract number DE-FG02-87ER40328 at the University of Minnesota, and would also like to thank SLAC for their hospitality and support during the period of completion of this project. IPG is supported by NSF Physics Frontier Center Award number 2020275. PM thanks the Neutrino Theory Network Program Grant under award number DE-AC02-07CHI11359, which supported this work while she was at UC Berkeley. Additionally, PM is grateful for the support of the Cambridge-Infosys AI Center for funding her current research.

\bibliographystyle{bibi}
\bibliography{Nucleo_bib}

\include{SMmod.tex}

\end{document}

%% file: SMmod.tex
\onecolumngrid
\appendix

\setcounter{equation}{0}
\setcounter{figure}{0}
\setcounter{table}{0}
\setcounter{page}{1}
\makeatletter
\renewcommand{\theequation}{S\arabic{equation}}
\renewcommand{\thefigure}{S\arabic{figure}}
\renewcommand{\thepage}{S\arabic{page}}
\renewcommand{\thetable}{S\arabic{table}}

\begin{center}
\textbf{\large Supplemental Material for the Letter\\[0.5ex]
{\em Solar-System Abundances of $p$-Nuclides Probe Collective Neutrino Oscillations in Supernovae}}
\end{center}

\bigskip
In this Supplemental Material (SM), we describe the methods employed in modeling the hydrodynamics, nucleosynthesis, and  flavor equilibration prescription. We present in detail the model parameters extracted from the {\tt GARCHING} supernova simulation, define the production factors used to assess the agreement with solar abundances, and extend our analysis in the main text to mildly supersonic models.
\bigskip

\section{A. Methods: hydrodynamics and nucleosynthesis}

We compute nucleosynthetic yields following a two-step approach, akin to Refs.~\cite{Friedland:2023kqp, Friedland:2025lge}. First, we construct the trajectories of tracer particles for a number of time
snapshots. Then, we obtain the yields by post-processing each tracer trajectory with the open-source nuclear reaction network {\tt SkyNet}~\cite{Lippuner:2017tyn}.

Tracer trajectories are divided into two phases. In the first, described as a steady-state neutrino-driven outflow (NDO), material accelerates from the protoneutron star (PNS) surface and subsequently decelerates through interaction with the surrounding ejecta.  In the second phase, the material joins the homologous expansion behind the forward shock and co-moves with it.

Whether the outflow becomes supersonic and develops a termination shock, or instead remains subsonic throughout, is not known a priori. The system often lies close to criticality~\cite{Friedland:2020ecy}, as our benchmark model demonstrates,
with the character of the outflow determined by the interplay of neutrino heating, PNS properties, and the confining pressure imposed by the progenitor profile.
Rather than postulating a supersonic or subsonic solution, we solve the hydrodynamic equations as a boundary value problem consistent with the confining pressure, at the edge of the hot bubble, inferred from the progenitor profile and front-shock velocity at each temporal snapshot. This method is described in detail in Ref.~\cite{Friedland:2025lge}. To incorporate the effect of neutrino oscillations, we posit a radius after which the neutrino flavors \textit{equilibrate} (see SM~C), which maximizes the effect of neutrino mixing, and vary this radius as a parameter. Concretely, this means that the neutrino spectra which enter the neutrino capture terms in the outflow calculation as well as in {\tt SkyNet}, are discretely modified before and after this mixing radius. This rather minimal prescription of positing a radius of equilibration does offer us a straightforward way of studying the radial dependence of oscillations' impact on nucleosynthesis agnostic of the microphysics.

The main ingredients of our model for the outflow have been delineated in the Letter and detailed in \cite{Friedland:2025lge}. Here we accentuate a few aspects. Although our hydrodynamic equations assume the steady-state condition, we consider the time evolution of the boundary conditions---including PNS radius, luminosities, confining pressure---and compute the outflow every $0.1$ s. This yields a series of snapshots. Our Equation of State (EoS) incorporates variable radiation degrees of freedom and the monatomic baryon gas. 
A key ingredient in determining the NDO is the neutrino net heating rate $\dot{Q}$, computed as in Ref.~\cite{Friedland:2025lge}. We include (i) charged-current absorption and emission on free nucleons, $\nu_e + n \leftrightarrow p + e^-$ and $\bar{\nu}_e + p \leftrightarrow n + e^+$, (ii) elastic neutrino–electron scattering $e + \nu \rightarrow e + \nu$, and (iii) cooling from electron-positron annihilation $e^- + e^+ \rightarrow \nu + \bar{\nu}$, following the prescription in Refs.~\cite{Qian:1996xt,Otsuki:1999kb}. Charged current absorptions, with a heating rate scaling with the third moment of the neutrino energy distribution---two powers from the cross section and one from the deposited energy---are the main heating processes and also determine the neutron-to-proton ratio in the outflow. Since SN simulations typically provide only the first two energy moments, we parametrize neutrino spectra (see SM~B) to reconstruct higher-order moments required to compute $\dot{Q}$. 

Nucleosynthesis is computed by post-processing each tracer with {\tt SkyNet}, using nuclear reaction rates from the {\tt Reaclib} library, version 2.2~\cite{reaclib2010},  with the modifications described in Ref.~\cite{Friedland:2025lge} (see Sec.~8 and App.~G therein), including for instance time-evolving neutrino luminosities, the blueshift of neutrino temperature and luminosities, medium-enhanced triple-$\alpha$ reaction rates from Refs.~\cite{Beard:2017jpg, Jin:2020}, and the correct decay rate for $^{92}$Nb~\cite{Friedland:2023kqp}. Additionally, for this analysis we further modify the code to account for neutrino flavor conversion at a mixing radius $r_{\rm mix}=R_{\rm PNS}(t)+\Delta r_{\rm mix}$. For each tracer, we start nucleosynthesis runs at $T=4.5$\,MeV, using as {\tt SkyNet} inputs the time evolution of its temperature $T$, density $\rho$, and its location $r$, along with a constant PNS mass and time-dependent PNS radius $R_{\rm PNS}(t)$, assumed to be equal to the neutrinosphere radius. The time evolution of the electron fraction $Y_e$ is computed by {\tt SkyNet} and it does incorporate weak-magnetism (WM) and recoil corrections with the method described in Ref.~\cite{Friedland:2025lge}.

In the absence of mixing, $Y_e$ approaches the asymptotic value given by~\cite{Horowitz:1999fe,Horowitz:2001xf,Burrows:2002jv}
\begin{equation}\label{eq:ye-wm}
Y_{e} \approx \left(1 + \frac{L_{\bar{\nu}_e}}{L_{\nu_e}} \frac{\epsilon_{\bar{\nu}_e}- 2\Delta - 7.1 {\varepsilon_{\bar\nu_e}^2}/{m_N} }{\epsilon_{\nu_e}+2\Delta + 1.1 {\varepsilon_{\nu_e}^2}/{m_N}} \right)^{-1},
\end{equation}
where $m_N$ is the nucleon mass, $\Delta=1.293$ MeV, $\epsilon_{\nu_i}=\langle E_{\nu_i}^2\rangle/\langle E_{\nu_i}\rangle$, and $\varepsilon_{\nu_i}=(\langle E_{\nu_i}^3\rangle/\langle E_{\nu_i}\rangle)^{1/2}$, with $\langle E_{\nu_i}^n\rangle$ the n-th moment of the energy distribution. {Note that this expression is derived using an approximation to the exact cross section, only including terms up to first order in $\Delta/E_{\nu_i}$ and $E_{\nu_i}/m_N$~\cite{Burrows:2002jv}. For a more exact treatment, and how it reduces to this expression at leading order, see Eqs. (22)--(25) from Ref.~\cite{Horowitz:2001xf}}. In presence of flavor conversion, the asymptotic value of $Y_e$ increases and depends on the radius at which conversion occurs, as discussed in the main Letter.

To incorporate flavor mixing, we modify {\tt SkyNet} to include $\nu_x$ and $\bar{\nu}_x$ in addition to $\nu_e$ and $\bar{\nu}_e$, which were the only species considered in our previous analyses. As input, we provide time-dependent unmixed neutrino luminosities and spectra. Each (anti)neutrino spectrum is modeled using a fit that enables the reconstruction of higher-order energy moments from those provided by supernova simulations (see SM~B). Starting from the original spectra of $\nu_e$, $\bar{\nu}_e$, $\nu_x$, and $\bar{\nu}_x$, we compute distance-dependent spectra and determine the corresponding mixed quantities at the radius $r_{\rm mix}$, as described in SM~C.

\section{B. Model parameters from simulations}

\begin{figure}[t!]
    \centering
\includegraphics[width=0.5\columnwidth]{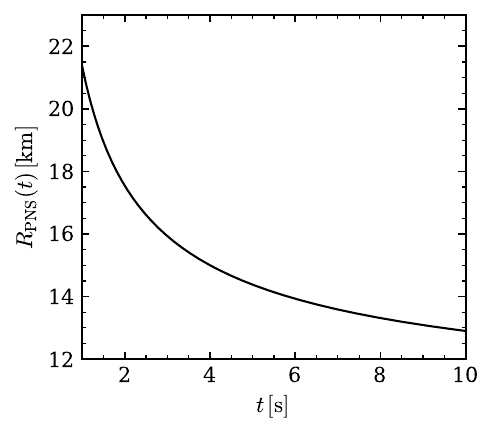}
    \caption{
   Evolution of the PNS radius as a function of post-bounce time for our benchmark $1.93\,M_\odot$ model, parametrized according to Eq.~(\ref{eq:PNSrad_abc}).}
    
    \label{fig:R_PNS}
\end{figure}

We model the NDO using input quantities extracted from simulations performed with the {\tt PROMETHEUS-VERTEX} neutrino-hydrodynamics code~\cite{Rampp:2002bq}. This code solves the fully energy- and velocity-dependent neutrino transport equations for all six species of neutrinos and antineutrinos, employing a state-of-the-art treatment of neutrino interactions~\cite{Buras:2005rp,Janka:2012wk,Bollig:2017lki}. As discussed in Ref.~\cite{Fiorillo:2023frv}, these simulations include a one-dimensional treatment of PNS convection through a mixing-length description of the convective fluxes~\cite{Mirizzi:2015eza}. They also account for the presence of muons in the hot PNS and the associated muonic-neutrino interactions~\cite{Bollig:2017lki}. However, the resulting differences between the luminosities and spectra of $\mu$ and $\tau$ neutrinos remain relatively small throughout most of the PNS cooling phase.

For our benchmark model, we adopt parameters from a 1D SN simulation from a $20\,M_\odot$ progenitor~\cite{Sukhbold:2015} performed by the {\tt GARCHING} collaboration~\cite{SNarchive}. This model possesses the qualitative features required for efficient $\nu p$ nucleosynthesis~\cite{Friedland:2023kqp}, specifically a massive PNS (with baryonic mass of $M=1.93\,M_\odot$) and subsonic outflows. We notice that such heavy PNS masses naturally arise in multi-dimensional simulations with massive progenitors~\cite{Bollig:2020phc,Janka:2025tvf}. 

In this model, the evolution of the PNS radius (solid line in Fig.~\ref{fig:R_PNS}) is well fitted by
\begin{equation}\label{eq:PNSrad_abc}
R_{\rm PNS}(t) = \left[a + b\,
\left(\frac{t}{1\, {\rm s}}\right)^{-c}\right]\,{\rm km}\,,
\end{equation}
where $t$ denotes the post-bounce time in seconds and the coefficients are $a=9.97$, $b=11.40$ and $c=0.59$. We adopt this as the near boundary condition for our outflow calculation. The far boundary conditions are fixed in our model by taking a front-shock (FS) velocity of $v_{\rm FS}=6000\,$km/s, consistent with the simulation results.

To assess the impact of an early transition from subsonic to supersonic outflows, we also perturb our benchmark model by considering a slightly lighter progenitor of $18.6\,M_\odot$~\cite{Sukhbold:2015}, with $v_{\rm FS}=7000\,$km/s, while keeping the PNS properties fixed. A more detailed discussion about this model is reported in SM~E.

Our reference SN simulation employs the SFHo EoS~\cite{Steiner+2013,Hempel+2010} and corresponds to the 1.93-SFHo model in Refs.~\cite{Fiorillo:2023frv,Lucente:2024ngp}. We describe below its main features.

The presence of convection accelerates PNS cooling and modifies the time evolution of the neutrino luminosity. As shown in Ref.~\cite{Roberts:2011yw}, a kink in the neutrino light curve when plotted in a doubly logarithmic scale marks the end of convection in the mantle region of the PNS, while convection remains active in the deeper high-density core. This feature typically occurs at $t \lesssim 10\,{\rm s}$~\cite{Pascal:2022qeg,Fiorillo:2023frv,Lucente:2024ngp}. Consequently, the neutrino luminosity in the cooling phase cannot be accurately described by a simple exponential or power-law decay when convection is present.

Following Ref.~\cite{Lucente:2024ngp}, we parametrize the neutrino luminosity as
\begin{equation}
L_{\nu}(t) = C_\nu\, t^{-\alpha_{\nu}}\,\exp\!\left[-\left(t/\tau_\nu\right)^{-n_\nu}\right],
\label{eq:lum}
\end{equation}
for each (anti)neutrino flavor $\nu = \nu_e,\,\bar{\nu}_e,\,\nu_x,\,\bar{\nu}_x$, with $x$ representing non-electron species. The parameters used in this work are taken from Tabs.~V and VI of Ref.~\cite{Lucente:2024ngp} for electron and non-electron (anti)neutrinos, respectively. We mention here that Tab.~VI in Ref.~\cite{Lucente:2024ngp} reports the parameter values for $\nu_\mu$. However, as discussed in Ref.~\cite{Lucente:2024ngp}, the differences between $\nu_\mu$ and $\nu_\tau$ are small, and we therefore use $\nu_\mu$ and $\bar{\nu}_\mu$ as representative of $\nu_x$ and $\bar{\nu}_x$.

To simplify the implementation of the time dependence of $L_\nu$ in the nuclear-reaction network {\tt SkyNet}~\cite{Lippuner:2017tyn}, we assume a common time evolution for all (anti)neutrino species. Specifically, the parameters $\alpha_\nu$, $\tau_\nu$, and $n_\nu$ are taken as the averages over all species. With this prescription, the optimal production window for the $\nu p$ process occurs at $t \simeq 3.5\,{\rm s}$ in our benchmark model. We therefore fix the normalization constants $C_\nu$ in order to reproduce the luminosities of all species at $t=3.5\,{\rm s}$. The fit parameters for $L_\nu$ in the $1.93\,M_\odot$ model are listed in Tab.~\ref{tab:nupar1.93}.

\begin{figure}[t!]
    \centering
\includegraphics[width=0.49\columnwidth]{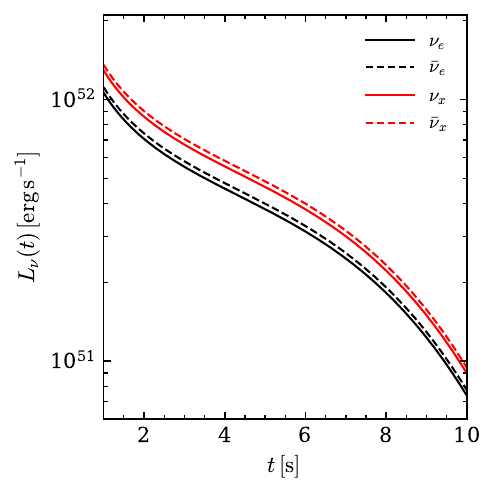}
\includegraphics[width=0.49\columnwidth]{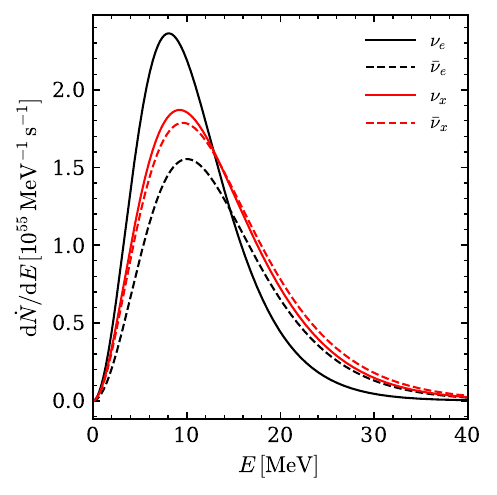}
    \caption{Time evolution of the neutrino luminosities (left panel) and production spectra at $t=3.5\,{\rm s}$ (right panel) for $\nu_e$ (solid black), $\bar{\nu}_e$ (dashed black), $\nu_x$ (solid red), $\bar{\nu}_x$ (dashed red), in our $1.93\,{M_\odot}$ model.}
    \label{fig:nu_quant_1.93}
\end{figure}

The resulting time evolution of $L_\nu(t)$ in our model is shown in the left panel of Fig.~\ref{fig:nu_quant_1.93}, exhibiting the hierarchy $L_{\nu_e} < L_{\bar{\nu}_e} < L_{\nu_x} < L_{\bar{\nu}_x}$ at any given time. The small difference between $L_{\nu_x}$ and $L_{\bar{\nu}_x}$ is mainly related to WM effects~\cite{Horowitz:2001xf}, which increase neutrino interaction rates and lower the rates for antineutrinos~(see, e.g., the discussion in Ref.~\cite{Keil:2002in}). These effects also lead to small differences between the $\nu_x$ and $\bar{\nu}_x$ spectra, as discussed below.

Following the default option in {\tt SkyNet}, in this analysis we model neutrino spectra with a normalized ``pinched'' Fermi-Dirac distribution~\cite{1989A&A...224...49J,1989A&AS...78..375J}
\begin{equation}
f_{\nu_i}(E) = \frac{N_{\nu_i}}{e^{E/T_{\nu_i}-\eta_{\nu_i}}+1},
\label{eq:FDdist}
\end{equation}
where the spectral parameters $T_{\nu_i}$ and $\eta_{\nu_i}$ are obtained directly from the same SN simulation, and $N_{\nu_i}$ is fixed by the luminosity. This parametrization provides an accurate fit to simulated spectra~\cite{Keil:2002in} and enables the reconstruction of higher-order energy moments from the moments supplied by the simulations.

Since SN simulations provide only the first $\langle E_{\nu_i} \rangle $ and second moments $\langle E_{\nu_i}^2 \rangle $  of the energy distributions, we relate the effective neutrino temperature $T_{\nu_i}$ and the degeneracy parameter $\eta_{\nu_i}$ to these quantities as
\begin{equation}
\langle E_{\nu_i}\rangle = T_{\nu_i}\,\frac{F_3(\eta_{\nu_i})}{F_2(\eta_{\nu_i})}\,,\qquad
\frac{\langle E_{\nu_i}^2\rangle}{\langle E_{\nu_i}\rangle} = T_{\nu_i}\,\frac{F_4(\eta_{\nu_i})}{F_3(\eta_{\nu_i})}\,,
\end{equation}
where $F_n(\eta_{\nu_i})$ are the Fermi integrals defined as
\begin{equation}
F_n(\eta_{\nu_i}) \equiv \int_0^\infty \frac{x^n}{e^{x-\eta_{\nu_i}}+1} dx\,= -\Gamma(n+1){\rm Li}_{n+1}(-e^{\eta_{\nu_i}})\,,
\label{eq:FDint}
\end{equation}
where ${\rm Li}_{n}$ is the polylogarithm of order $n$.

Since $\langle E_{\nu_i} \rangle$ and $\langle E_{\nu_i}^2 \rangle$ vary only mildly during the first $\sim 5\,\mathrm{s}$ after core bounce (see, e.g., Fig.~1 of Ref.~\cite{Fiorillo:2023frv}), we use time-independent spectral parameters $T_{\nu_i}$ and $\eta_{\nu_i}$. These are fixed by matching $\langle E_{\nu_i} \rangle$ and $\langle E_{\nu_i}^2 \rangle$ to the values extracted from the simulations at the reference time $t=3.5\,\mathrm{s}$. The adopted values of $\langle E_{\nu_i} \rangle$, $\langle E_{\nu_i}^2 \rangle$, $T_{\nu_i}$, and $\eta_{\nu_i}$ for our benchmark model are reported in Tab.~\ref{tab:nupar1.93}.

The right panel of Fig.~\ref{fig:nu_quant_1.93} shows the production spectra (in MeV$^{-1}$ s$^{-1}$) for electron (black) and non-electron (red) neutrinos (solid lines) and antineutrinos (dashed lines) at $t=3.5\,{\rm s}$, normalized to reproduce the corresponding luminosities at that time. For each flavor, neutrino spectra peak at lower energies than the corresponding antineutrino spectra, with $\langle E_{\bar{\nu}_e} \rangle$ about $25\%$ larger than $\langle E_{\nu_e} \rangle$ and $\langle E_{\bar{\nu}_x} \rangle$ about $4\%$ larger than $\langle E_{\nu_x} \rangle$. The few-percent difference between the $\nu_x$ and $\bar{\nu}_x$ spectra arises from WM effects; in their absence the two spectra would be essentially identical.

With the assumed values of neutrino luminosity and spectra, the equilibrium value of  $Y_e$ in the NDO (see Eq.~\eqref{eq:ye-wm}) is approximately $0.57$.

\begin{table}[t!]
\centering
\begin{tabular}{lcccc}
\hline\hline
 & $\nu_e$ & $\bar{\nu}_e$ & $\nu_x$ & $\bar{\nu}_x$ \\
\hline
$C_\nu\,[10^{51}\,\mathrm{erg/s}]$ & 10.61 & 11.14 & 12.90 & 13.56 \\
$\alpha_\nu$ & 0.584 & 0.584 & 0.584 & 0.584 \\
$\tau_\nu\,\,\,[\mathrm{s}]$ & 9.31 & 9.31 & 9.31 & 9.31 \\
$n_\nu$ & 3.95 & 3.95 & 3.95 & 3.95 \\
\hline
$\langle E_\nu \rangle\,[\mathrm{MeV}]$ & 10.80 & 13.58 & 13.10 & 13.70 \\
$\langle E_\nu^2 \rangle\,[\mathrm{MeV}^2]$ & 148.03 & 235.25 & 223.34 & 244.98 \\
$T_\nu\,\,\,\,\,\,\,[\mathrm{MeV}]$ & 3.21 & 4.09 & 4.14 & 4.36 \\
$\eta_\nu$ & 1.17 & 0.97 & 0.06 & $-0.12$ \\
\hline\hline
\end{tabular}
\caption{Input parameters for neutrino luminosities (upper blocks) and energy spectra (lower blocks) of all (anti)neutrino species used for our benchmark $1.93\,M_\odot$ model.}
\label{tab:nupar1.93}
\end{table}

\section{C. Mixing of neutrino flavors}
\label{app:nu_mix}

As a benchmark for the effect of collective neutrino oscillations, we consider complete flavor equilibration, which transforms distributions as 
\begin{equation}
f'_{\nu_e}=f'_{\nu_x}=\frac{f_{\nu_e}+2f_{\nu_x}}{3}, \qquad
f'_{\bar{\nu}_e}=f'_{\bar{\nu}_x}=\frac{f_{\bar\nu_e}+2f_{\bar\nu_x}}{3}\,,
\label{eq:equilibrate}
\end{equation}
where, under the assumption of flavor equipartition, the mixed (anti)neutrino distribution function receives equal contributions from $\nu_e$ ($\bar{\nu}_e$) and the individual heavy-lepton (anti)neutrinos. As discussed in the main text, our results remain valid provided that oscillations convert a substantial fraction of the heavy-lepton-flavor (anti)neutrino flux into electron (anti)neutrinos.
Assuming the pre-mixing spectra are of Fermi-Dirac shape as reported in Eq.~\eqref{eq:FDdist}, the mixed luminosities are

\begin{equation}
L'_{\nu_e} = L'_{\nu_x} = \frac{L_{\nu_e}+2L_{\nu_x}}{3}, \qquad
L'_{\bar\nu_e} = L'_{\bar\nu_x} = \frac{L_{\bar\nu_e}+2L_{\bar\nu_x}}{3}.
\end{equation}

and the energy moments are

\begin{equation}\label{eq:En}
\langle E_{\nu_e}^{n} \rangle' = \langle E_{\nu_x}^{n} \rangle'
= \frac{N_{\nu_e} T_{\nu_e}^{n+3} F_{n+2}(\eta_{\nu_e}) + 2\, N_{\nu_x} T_{\nu_x}^{n+3} F_{n+2}(\eta_{\nu_x})}
       {N_{\nu_e} T_{\nu_e}^{3} F_{2}(\eta_{\nu_e}) + 2\, N_{\nu_x} T_{\nu_x}^{3} F_{2}(\eta_{\nu_x})},
\end{equation}
with $F_n$ as the Fermi-Dirac integral of power $n$ defined in Eq.~\eqref{eq:FDint}. An analogous expression for $\langle E_{\bar\nu_e}^{n} \rangle' = \langle E_{\bar\nu_x}^{n} \rangle'$ follows by replacing $\nu_\alpha \to \bar{\nu}_\alpha$ ($\alpha=e,x$) everywhere in Eq.~\eqref{eq:En}.

\begin{figure}[t!]
    \centering
\includegraphics[width=0.49\columnwidth]{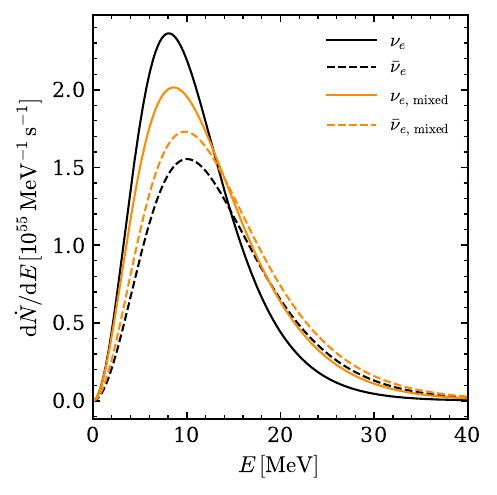}
\includegraphics[width=0.49\columnwidth]{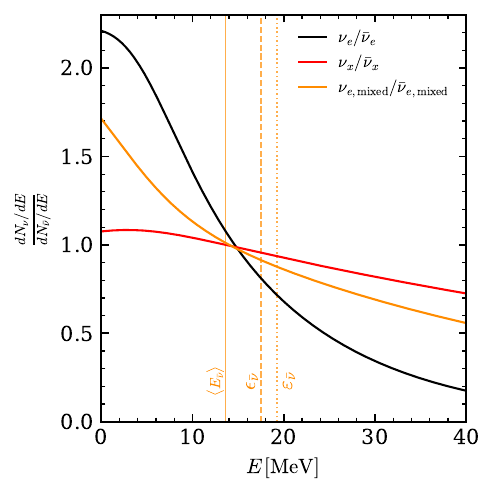}
    \caption{\emph{Left:} Production spectra at $t=3.5\,{\rm s}$ for unmixed (black) and mixed (orange) electron neutrinos (solid curves) and antineutrinos (dashed curves), in our $1.93\,{M_\odot}$ model. \emph{Right:} Ratio of neutrino to antineutrino spectra for the unmixed electron flavor (black) and non-electron flavors (red), as well as for the electron flavor in the mixed case (orange) in the $1.93\,M_\odot$ model. The vertical orange lines correspond to the average energy $\langle E\rangle$ (solid) and its inverse ratio with the second $\epsilon=\langle E^2\rangle/\langle E \rangle$ (dashed) and third $\varepsilon = \sqrt{\langle E^3 \rangle/\langle E \rangle}$ (dotted) moments of the mixed electron antineutrino's energy distribution.}
    \label{fig:spec_mix_1.93}
\end{figure}

The left panel of Fig.~\ref{fig:spec_mix_1.93} shows the unmixed (black) and mixed (orange) production spectra of electron neutrinos (solid) and antineutrinos (dashed) at $t=3.5\,{\rm s}$. The contribution of non-electron (anti)neutrinos---whose spectra are more similar (see the right panel of Fig.~\ref{fig:nu_quant_1.93})---reduces the difference between neutrino and antineutrino spectra after mixing. The resulting mixed luminosities and spectra yield $Y_e \approx 0.6$ via Eq.~\eqref{eq:ye-wm} and are used in the outflow calculation described in SM~A.

A more quantitative comparison is obtained by considering the ratio of neutrino to antineutrino spectra, shown in the right panel of Fig.~\ref{fig:spec_mix_1.93} for the unmixed electron flavor (black), non-electron flavors (red), and the mixed electron flavor (orange). In the mixed case, the spectrum receives a $1/3$ contribution from electron flavors and $2/3$ from non-electron species, so the ratio lies between the unmixed cases. Vertical lines indicate the average energy $\langle E_\nu \rangle$ (solid), as well as the parameters $\epsilon_\nu$ (dashed) and $\varepsilon_\nu$ (dotted) for the mixed antineutrino spectra. These quantities enter the determination of $\dot{Q}$ and $Y_e$ in Eq.~\eqref{eq:ye-wm}. Due to the similarity between $\nu_x$ and $\bar{\nu}_x$ spectra, the ratio of mixed $\nu_e$ to $\bar{\nu}_e$ spectra is close to unity near $\epsilon$ and $\varepsilon$, explaining why $Y_e \approx 0.6$ is obtained.

\begin{figure}[t!]
    \centering
\includegraphics[width=0.49\columnwidth]{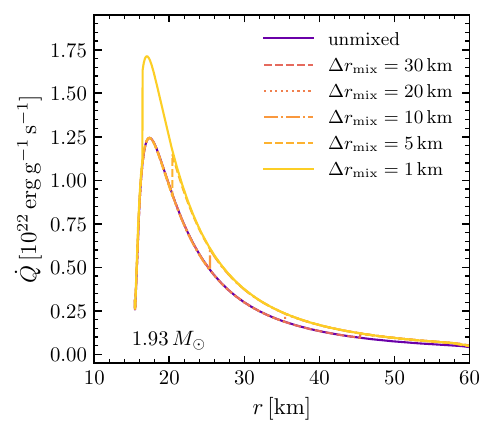}
\includegraphics[width=0.49\columnwidth]{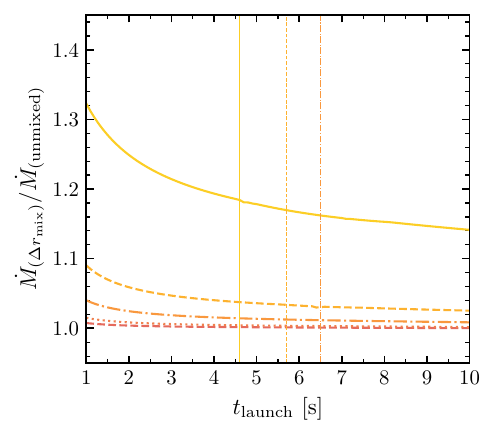}
    \caption{\emph{Left:} Neutrino net heating rate $\dot{Q}$ as a function of radius $r$ at $t=3.5$\,s for flavor conversion at different distances $\Delta r_{\rm mix}$ from the PNS surface. \emph{Right:} Relative enhancement of the mass outflow rate $\dot{M}$ as a function of launch time $t_{\rm launch}$ for different mixing radii $\Delta r_{\rm mix}$. Vertical lines denote the beginning of the transonic case for $\Delta r_{\rm mix} = 1$\,km (solid), 5\,km (dashed) and 10\,km (dash-dotted). Color coding follows Fig.~\ref{fig:Integrated_yields}.}
    \label{fig:Qdot}
\end{figure}

Larger values of $\varepsilon_\nu$ enhance the neutrino net heating rate $\dot{Q}$ and consequently $\dot{M}$. The left panel of Fig.~\ref{fig:Qdot} shows the radial profile of $\dot{Q}$ at $t=3.5$\,s near the PNS for our benchmark model, for flavor mixing at different distances $\Delta r_{\rm mix}$ from the PNS surface. For each mixing scenario, deviations from the unmixed case are negligible at $r<R_{\rm PNS}+\Delta r_{\rm mix}$, while at larger radii $\dot{Q}$ increases by $\sim 30\%$ due to higher neutrino luminosities and harder spectra. Enhanced $\dot{Q}$ closer to the PNS accelerates the outflow and increases $\dot{M}$.

The right panel of Fig.~\ref{fig:Qdot} shows the time evolution of $\dot{M}$ enhancement relative to the unmixed case for different mixing radii $\Delta r_{\rm mix}$. The relative enhancement decreases with time: it is $\lesssim 30\%$ for $\Delta r_{\rm mix}=1$\,km, $\lesssim 10\%$ for $\Delta r_{\rm mix}=5$\,km, and $<5\%$ for larger mixing radii. The increased $\dot{Q}$ also raises the outflow velocity, inducing a transonic transition at $\Delta r_{\rm mix}=1$\,km (vertical solid line), 5\,km (dashed line), and 10\,km (dash-dotted line). For $\Delta r_{\rm mix}\lesssim 10$\,km, the transonic transition occurs at progressively later times as $\Delta r_{\rm mix}$ increases. For $\Delta r_{\rm mix}\gtrsim 10$\,km, the transonic transition disappears and the velocity enhancement remains below 10\% relative to the unmixed case.

\section{D. Integrated yields and production factors}

To compare $\nu p$ yields with solar-system abundances~\cite{Lodders:2003}, we compute time-integrated yields. For a nuclide $(A,Z)$, the time-averaged abundance is defined as~\cite{Friedland:2023kqp}
\begin{equation}
    \langle Y_{A,\,Z}\rangle = \frac{\int dt\, Y_{A,Z}(t)\dot{M}(t)}{\int dt\, \dot{M}(t)}\,,
\end{equation}
where $Y_{A,Z}(t)$ and $\dot{M}(t)$ denote the instantaneous abundance and mass outflow rate, respectively. The resulting time-averaged yields $\langle Y_A \rangle = \sum_Z Y_{A,Z}$ are shown for the benchmark $20\,M_\odot$ model in Fig.~\ref{fig:Integrated_yields} and for the $18.6\,M_\odot$ model in the left panel of Fig.~\ref{fig:quantities_18.6}. 

\begin{figure}[!]
    \centering
    \includegraphics[width=0.49\columnwidth]{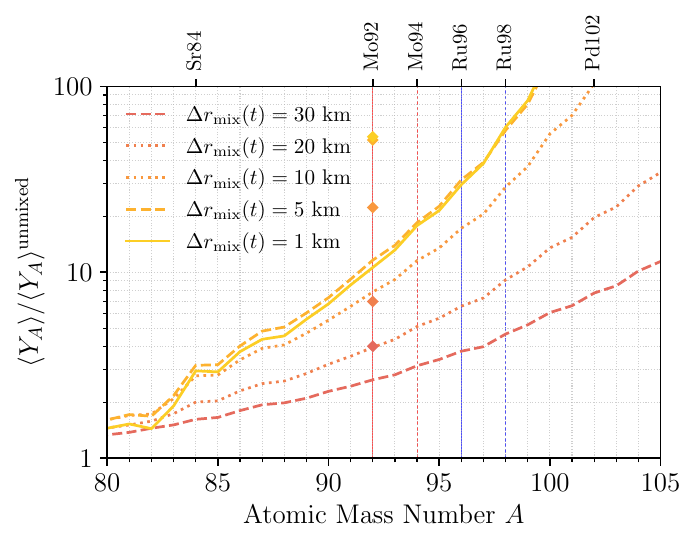}
        \includegraphics[width=0.49\columnwidth]{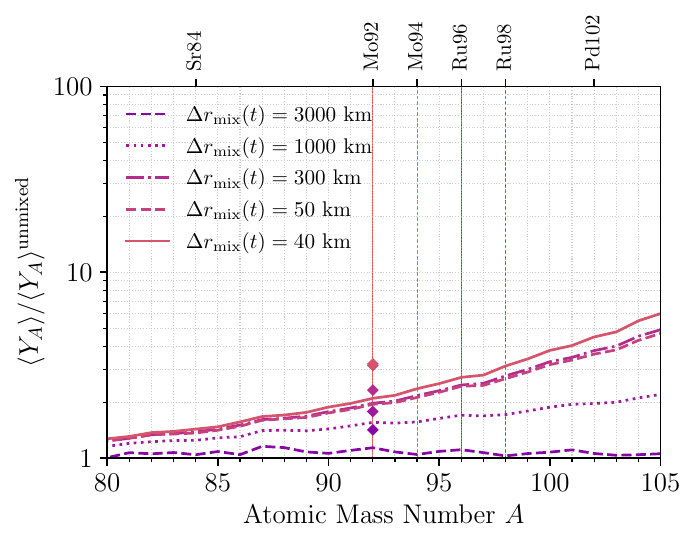}
        \caption{Enhancement factors $\langle Y_A \rangle/\langle Y_A\rangle^{\rm unmixed}$ for time-averaged yields under flavor equilibration at various radii, for near (left) and far (right) conversions. Curves show enhancement versus mass number $A$; diamonds mark $^{92}$Nb enhancement. Vertical lines denote $^{92}$Mo (solid red), $^{94}$Mo (dashed red), $^{96}$Ru (solid blue), and $^{98}$Ru (dashed blue). Color coding follows Fig.~\ref{fig:Integrated_yields}.}
    \label{fig:Yields_ratio}
\end{figure}

Time-averaged yields increase as mixing occurs closer to the PNS, as discussed in the main text. Figure~\ref{fig:Yields_ratio} shows enhancement factors $\langle Y_A\rangle/\langle Y_A\rangle^{\rm unmixed}$ for $p$-nuclides (curves) and $^{92}$Nb (diamonds) under near (left) and far (right) conversions. The enhancement grows with mass number $A$ in all scenarios.

For near conversions, the $\Delta r_{\rm mix}=1,5$~km cases nearly coincide, yielding factors of $\sim10$ for $^{92}$Mo and $\sim60$ for $^{98}$Ru. At $\Delta r_{\rm mix}=10$~km, enhancements remain within a factor of 2 of these values: $\sim8$ for $^{92}$Mo and $\sim30$ for $^{98}$Ru. The $\Delta r_{\rm mix}=20,30$~km cases produce smaller boosts: $\lesssim10$ and $\lesssim5$, respectively.

Far conversions yield more modest increases. For $\Delta r_{\rm mix}=40$--$300$~km, enhancement factors remain $\lesssim3$. At larger radii, where conversion occurs during the $p$-nuclide production phase, enhancement diminishes further (dotted purple line, $\Delta r_{\rm mix}=1000$~km) and becomes negligible for $\Delta r_{\rm mix}\gtrsim3000$~km (dashed violet line).

The $^{92}$Nb increase differs qualitatively. Near conversion enhances its abundance by up to 2 orders of magnitude, with factors of $\sim50$ for $\Delta r_{\rm mix}=1,5$~km. Far conversion produces a factor of $\sim3$ increase at $\Delta r_{\rm mix}=40,50$~km, and enhancement persists even beyond 1000~km, reaching 1.4 at $\Delta r_{\rm mix}=3000$~km (violet diamond).

From these abundances we define the isotopic production factor~\cite{Wanajo:2010mc,Friedland:2023kqp,Friedland:2025lge}
\begin{equation}
f_{A,Z} = \frac{\langle X_{A,Z} \rangle}{X_{A,Z}^\odot}\,,
\label{eq:fAZ}
\end{equation}
where $X_{A,Z}^\odot$ is the solar-system mass fraction of isotope $(A,Z)$, and $\langle X_{A,Z}\rangle = A\,\langle Y_{A,Z}\rangle$ is the time-averaged mass fraction obtained from the {\tt SkyNet} output, with $\sum_{A,Z} A\,Y_{A,Z}=1$. Solar abundances are taken from meteoritic measurements of carbonaceous chondrites~\cite{Lodders:2003}, following Ref.~\cite{Friedland:2025lge}.

Production factors quantify both relative and absolute abundances. Accounting for astrophysical uncertainties, nuclides are considered co-produced if $f_{A,Z} \gtrsim f_{\rm max}/10$, where $f_{\rm max}$ is the largest production factor in a given model~\cite{Wanajo:2010mc,Bliss:2018djg}. Absolute yields are characterized by the overproduction factor~\cite{Woosley:1994ux,Wanajo:2010mc,Friedland:2023kqp}
\begin{equation}
f^{\rm ov}_{A,Z} = f_{A,Z}\,\frac{M_{\rm out}}{M_{\rm ejec}}\,,
\label{eq:fov}
\end{equation}
where the ratio $(M_{\rm out}/M_{\rm ejec})$ is the astrophysical dilution factor, with $M_{\rm out}=\int dt   \,\dot{M}(t)$ the total mass driven out in the NDO and $M_{\rm ejec}\simeq M_{\rm prog}-M_{\rm PNS}$ the total mass ejected in the explosion. Values $f^{\rm ov}_{A,Z}\gtrsim 10$ are typically required to reproduce solar abundances~\cite{Woosley:1994ux,Wanajo:2010mc,Friedland:2023kqp}. The ejected mass $M_{\rm ejec}$ is fixed by the underlying SN simulation and is therefore independent of the mixing prescription, with $M_{\rm ejec}=18.07\,M_\odot$ for the $20\,M_\odot$ progenitor and $M_{\rm ejec}=16.67\,M_\odot$ for the $18.6\,M_\odot$ progenitor. Flavor mixing increases neutrino heating in the hot bubble and thus raises the mass-outflow rate $\dot M$. As a result, flavor equilibration occurring closer to the PNS yields a larger total outflow mass, $M_{\rm out}$, and a greater dilution factor. For the $20\,M_\odot$ model, we find $M_{\rm out}\simeq 3.5\times10^{-3}\,M_\odot$ (dilution factor $\simeq2.0\times10^{-4}$) without mixing, rising to $M_{\rm out}\simeq 4.4\times10^{-3}\,M_\odot$ (dilution factor $\simeq2.4\times10^{-4}$) for the $\Delta r_{\rm mix}=1$ km case. For the $18.6\,M_\odot$ model, we obtain $M_{\rm out}\simeq 3.6\times10^{-3}\,M_\odot$ (dilution factor $\simeq2.1\times10^{-4}$) in the unmixed case, and $M_{\rm out}\simeq4.4\times10^{-3}\,M_\odot$ (dilution factor $\simeq2.6\times10^{-4}$) for $\Delta r_{\rm mix}=1$ km.

\begin{figure}[!]
    \centering
    \includegraphics[width=0.49\columnwidth]{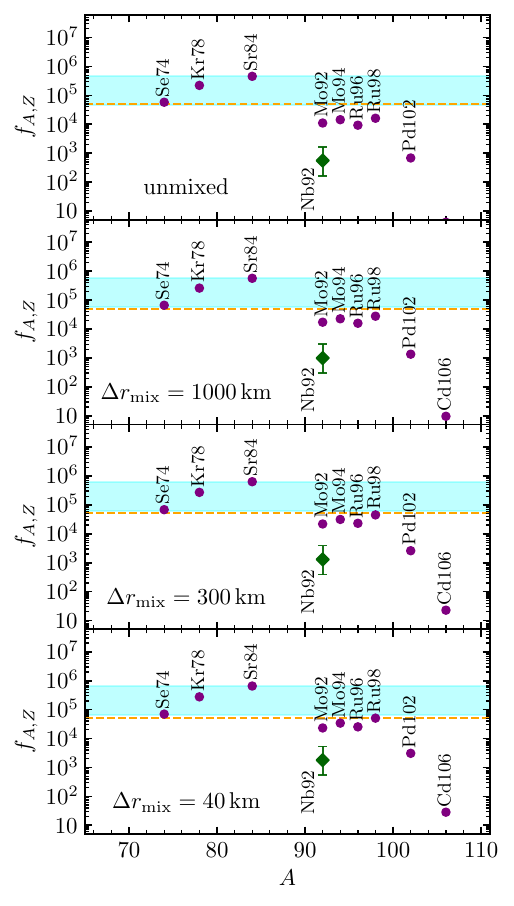}
        \includegraphics[width=0.49\columnwidth]{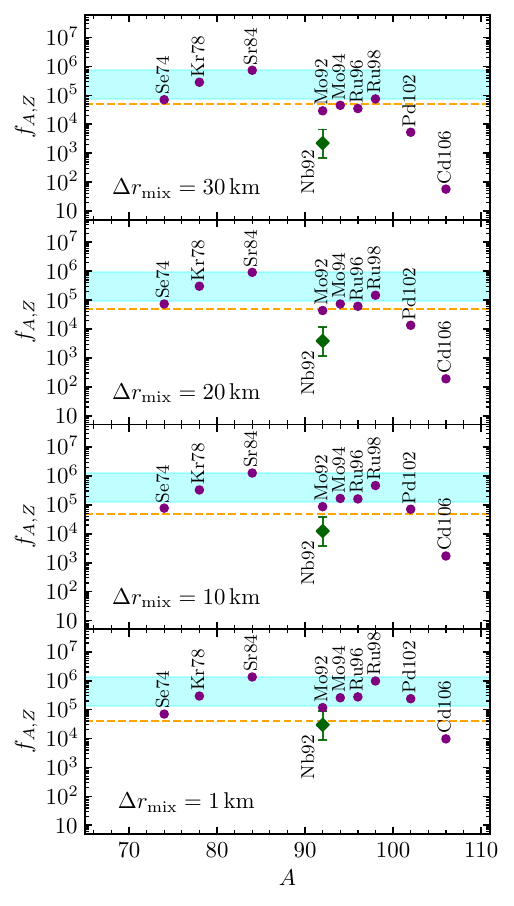}
    \caption{Production factors of $p$-nuclides (purple dots) and $^{92}{\rm Nb}$ (green diamond with error bars) in the $20\,M_\odot$ model, under flavor conversion occurring at different radii $\Delta r_{\rm mix}$ from the PNS. In each panel, the cyan band indicates the co-production region, and the dashed orange line marks the threshold required to satisfy solar observations. The uncertainty on $f(^{92}\mathrm{Nb})$ arises because $^{92}\mathrm{Nb}$ is an extinct radionuclide: its solar abundance cannot be measured directly, but is inferred from meteoritic data to correspond to a production fraction $^{92}\mathrm{Nb}/^{92}\mathrm{Mo}$ within $\sim10^{-3}$–$10^{-2}$.}
    \label{fig:Prod_factors}
\end{figure}

We show the production factors of different $p$-nuclides for our benchmark $20\,M_\odot$ model in Fig.~\ref{fig:Prod_factors} for mixing occurring at different distances $\Delta r_{\rm mix}$ from the PNS surface, comparing ``far'' conversion (left panel) and ``near'' conversion (right panel). Each panel also displays the production factor of the extinct radionuclide $^{92}{\rm Nb}$ (green diamond), normalized to $3\times10^{-3}$ of the $^{92}$Mo solar abundance. The associated error bars reflect the inferred production ratio $^{92}{\rm Nb}/^{92}{\rm Mo}$, constrained to lie in the range $\sim 10^{-3}$–$10^{-2}$ to account for the solar-system abundance of $^{92}{\rm Zr}$, into which $^{92}{\rm Nb}$ decays~\cite{Rauscher:2013,Lugaro:2016zuf,Iizuka:2016,Hibiya:2023}.

Production factors increase as mixing occurs closer to the PNS. For far conversion, Mo and Ru isotope production factors increase by more than a factor of 2 from the unmixed case to $\Delta r_{\rm mix}=40$\,km, while results at $\Delta r_{\rm mix}=300$\,km and 40\,km are the same. For near conversion, production factors increase monotonically from $\Delta r_{\rm mix}=30$\,km to 1\,km due to enhanced $Y_e$ and $\dot{M}$.

In all cases, $^{84}\mathrm{Sr}$ has the highest production factor. The co-production band (cyan) includes all isotopes with $f_{A,Z}\ge 0.1\,f(^{84}\mathrm{Sr})$. Without mixing, all species with $A\ge 90$ fall below this threshold. The increase induced by far conversion is not enough to reach the co-production region. For $\Delta r_{\rm mix}\lesssim 30$\,km, production factors approach co-production, and for $\Delta r_{\rm mix}\lesssim 10$\,km, isotopes up to $^{102}$Pd are co-produced.  For $\Delta r_{\rm mix}=1\,$km, flavor mixing increases Mo and Ru isotopes by a factor $\sim 10$, $^{92}\mathrm{Nb}$ by $\sim 100$, and enhances $^{102}\mathrm{Pd}$ and $^{106}\mathrm{Cd}$ by $\sim 2$ and more than $3$ orders of magnitude, respectively.

The horizontal dashed orange line marks the threshold $f_{A,Z}^{\rm ov}=10$ required to reproduce solar abundances [Eq.~\eqref{eq:fov}]. Mixing increases $\dot{M}$ through enhanced heating, yielding larger $M_{\rm out}$ and thus slightly lowering this threshold for smaller $\Delta r_{\rm mix}$. In the unmixed scenario, nuclides with $90\le A\le 100$ are neither co-produced nor sufficiently abundant to meet solar constraints. Far conversion cannot reach the solar threshold (all nuclides remain below the orange line), while Mo and Ru isotopes satisfy solar observations for $\Delta r_{\rm mix}\lesssim 30$\,km. For $\Delta r_{\rm mix}\lesssim 10$\,km, Mo and Ru isotopes---as well as $^{92}{\rm Nb}$ within uncertainties---are co-produced and meet solar-abundance requirements.

From the overproduction factor, we derive the solar-abundance thresholds shown in Fig.~\ref{fig:Integrated_yields} and Figs.~\ref{fig:quantities_18.6} by inverting the condition $f^{\rm ov}_{A,Z}\gtrsim 10$~\cite{Woosley:1994ux,Wanajo:2010mc,Friedland:2023kqp}:
\begin{equation}
\langle Y_A \rangle_{\rm thr} = 10\,\frac{M_{\rm ejec}}{M_{\rm out}}\,\frac{X_{A,Z}^\odot}{A}\,.
\end{equation}
For $^{92}$Nb, the minimum threshold corresponds to normalizing its production to $10^{-3}$ of the $^{92}$Mo abundance, while the black dot denotes the benchmark normalization of $3\times10^{-3}$.

To remain conservative, we always show $\langle Y_A \rangle_{\rm thr}$ only for the unmixed case, which yields the largest thresholds relative to the corresponding mixed scenarios.

\begin{figure}[t!]
    \centering  \includegraphics[width=0.7\columnwidth]{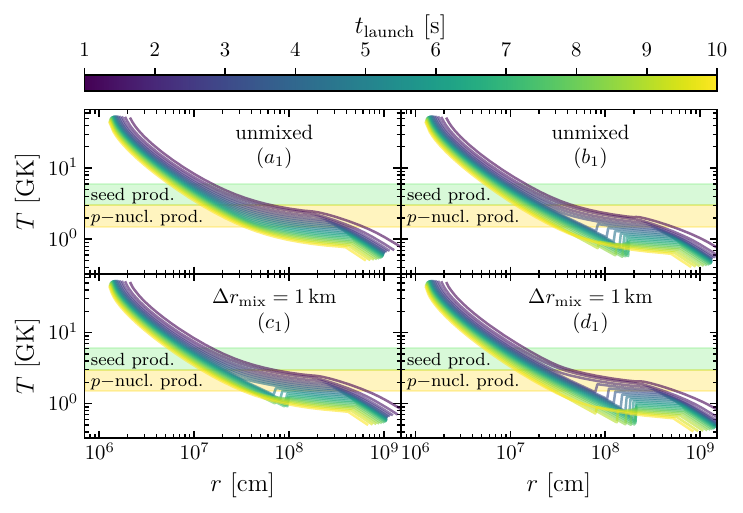}   \includegraphics[width=0.7\columnwidth]{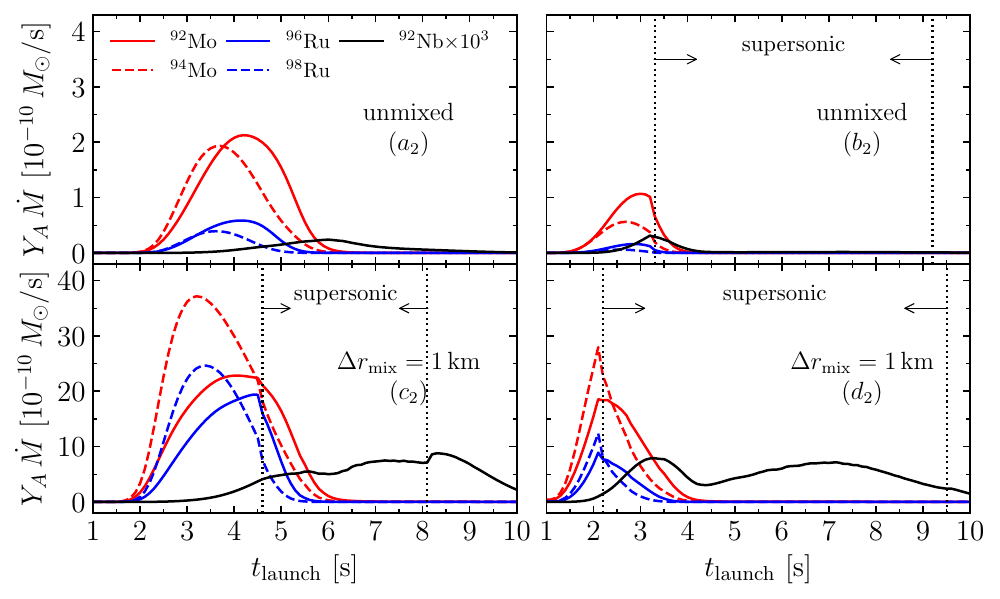}
    \caption{\emph{Upper panels:} Radial temperature profiles for outflows launched at $t_{\rm launch}\in[1,\,10]\,{\rm s}$, depicted here in steps of $0.5\,{\rm s}$. The green and yellow bands correspond to the seed-formation and $p$-nuclide formation stages of the $\nu p$-process, respectively~\cite{Wanajo:2010mc,Friedland:2025lge}. \emph{Upper panels:} Time evolution of the production rates of Mo (red) and Ru isotopes (blue), and $^{92}{\rm Nb}$ (black), with vertical dotted lines delimiting the time window where the outflows contain a supersonic transition. In each set of plots, the left (right) panels refer to the $20~M_\odot$ ($18.6~M_\odot$) progenitor, while the upper (lower) panels are dedicated to the unmixed ($\Delta r_{\rm mix}=1$~km) case. }
\label{fig:time_ev}
\end{figure}

\section{E. Mildly-supersonic models}

A natural question one has, after seeing the results of our benchmark $20\ M_\odot$ model, is whether the conditions of this simulation, albeit taken off the shelf, are ``fine-tuned" or coincidental so that when mixing is introduced, the supersonic transition enters only after the nucleosynthesis has mostly concluded. Had the transition occurred earlier and nucleosynthesis operated in a supersonic environment, there could be a real competition between the favorable influence of increased $Y_e$ and the adverse effect of the supersonic regime. 

To this end, we perturb the far boundary condition of our benchmark model by adopting a progenitor of $18.6\,M_\odot$ with $v_{\rm FS}=7000\ {\rm km/s}$, while retaining the PNS properties and neutrino spectra. This lowers the confining pressure of the hot bubble and increases the outflow's velocity, inducing a supersonic transition within the first seconds of the cooling phase without the introduction of neutrino oscillations. Considering such a mildly supersonic model affords us insights into the other side of near-critical supernova outflows.

Figure~\ref{fig:time_ev} compares the time evolution of the benchmark $20\,M_\odot$ model (left panels) and the perturbed $18.6\,M_\odot$ model (right panels) during the cooling phase ($t_{\rm launch}\in[1,10]\,$s). The upper panels show the evolution of the outflows in the unmixed (panels $a_1$ and $b_1$) and in the $\Delta r_{\rm mix}=1\,$km cases (panels $c_1$ and $d_1$), while the lower panels display the corresponding production rates as a function of the launch time. In contrast to the benchmark model, this $18.6\,M_\odot$ (panel $b_1$) model exhibits a transonic transition at $t_{\rm launch}\simeq 3.3\,$s, followed by a return to subsonic outflows at $t\simeq 9.2\,$s with the decline of neutrino luminosities. Flavor equilibration accelerates the onset of this transonic transition, which takes place at $t\simeq 2.2\,$s when mixing occurs $1\,$km above the PNS surface (panel $d_1$). This presents an interesting study as the induced supersonic regime now intrudes on the early cooling phase, when the $\nu p$ process is most efficient. 

What is the impact on nucleosynthesis? As shown in the lower panels of Fig.~\ref{fig:time_ev}, without mixing, the production rates in the optimal window (prior to the transonic transition at $t\simeq 3.3\,$s) are smaller by a factor of $\sim 2$ compared to the benchmark model (panel $a_2$). When mixing occurs $1\,$km above the PNS, the supersonic transition occurs near the end of the optimal production window in the $20\,M_\odot$ model (panel $c_2$), whereas in this $18.6\,M_\odot$ model (panel $d_2$) it clips the synthesis of $p$-nuclides, preventing the $\nu p$ process from attaining its peak rates.

Despite this hydrodynamic effect, in both models mixing at $\Delta r_{\rm mix}=1$\,km enhances the production rates by a factor of $\sim 20$ for Mo and $\sim 50$ for Ru isotopes during the optimal window. The $\sim 1\,$s reduction in the duration of this window is subdominant relative to this enhancement from the increased $Y_e$ and $\dot{M}$. Thus, in both the $20\, M_\odot$ and $18.6\, M_\odot$ models, effective neutrino mixing near the PNS renders higher nucleosynthetic yields. The production of $^{92}{\rm Nb}$ (black lines) is similarly improved in both models, with flavor equilibration at $\Delta r_{\rm mix}=1$\,km extending the production window up to $\sim 10\,$s.

\begin{figure*}[t!]
    \centering
\adjustbox{valign=c}{\includegraphics[width=0.49\columnwidth]{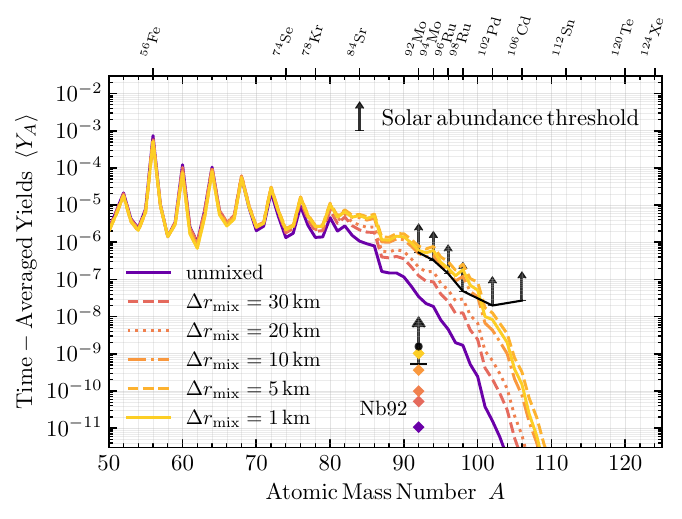}}
\adjustbox{valign=c}{\includegraphics[width=0.49\columnwidth]{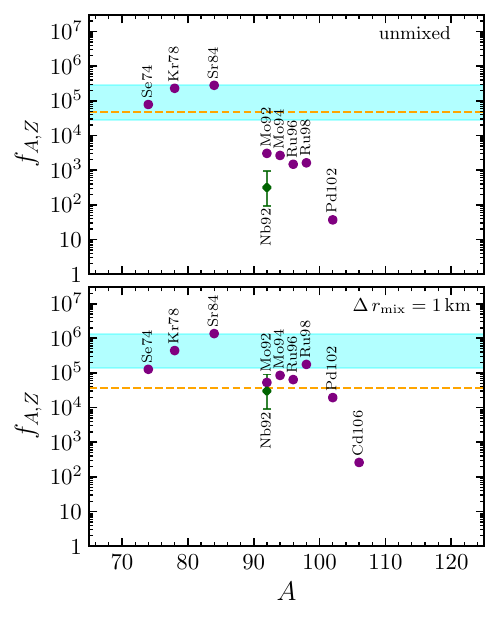}}
    \caption{\emph{Left:} Time-averaged yields $\langle Y_A\rangle$ for the $18.6\,M_\odot$ model under flavor conversion occurring at different radii from the PNS. Curves show $\langle Y_A\rangle$ versus mass number $A$; diamonds mark the corresponding $^{92}$Nb yield. Black arrows indicate the minimum abundances needed to reproduce the solar values. Color coding follows Fig.~\ref{fig:Integrated_yields}.
\emph{Right:} Production factors of $p$-nuclides in the $18.6\,M_\odot$ model, for the unmixed case (upper right panel) and flavor mixing occurring $1\,\mathrm{km}$ above the PNS ($\Delta r_{\rm mix}=1\,\mathrm{km}$---lower right panel). In each panel, the cyan band indicates the co-production region, and the dashed orange line marks the threshold corresponding to an overproduction factor of 10, required to account for solar abundances.}
    \label{fig:quantities_18.6}
\end{figure*}

The time-integrated results are reported in Fig.~\ref{fig:quantities_18.6}. Its left panel shows the time-averaged abundances for the $18.6\,M_\odot$ model, comparing the unmixed case (solid black line) with scenarios of varying radii of flavor equilibration, given by the displacement from the PNS radius $\Delta r_{\rm mix}$. The qualitative behavior is similar to that of the benchmark $20\,M_\odot$ model (Fig.~\ref{fig:Integrated_yields}): mixing magnifies $\nu p$ yields, generally as the equilibration radius approaches the PNS. For $\Delta r_{\rm mix}=20$--$30$\,km, yields increase by less than one order of magnitude, while when flavor conversion occurs sufficiently close to the PNS ($\Delta r_{\rm mix}\lesssim 10\,$km), $\nu p$ yields increase by up to a factor of $\sim 100$ relative to their original values. Notably, the $\Delta r_{\rm mix}=5$\,km case produces slightly larger yields than the $\Delta r_{\rm mix}=1$ km scenario. This occurs because the earlier transonic transition in the $\Delta r_{\rm mix}=1$\,km case partially counteracts the larger $Y_e$ boost, resulting in a net reduction compared to $\Delta r_{\rm mix}=5$\,km.
Overall, the yields from the $18.6\,M_\odot$ model are systematically lower by about one order of magnitude compared to the corresponding cases in the benchmark model. Consequently, the time-averaged yields of $^{92,\,94}{\rm Mo}$ and $^{96,\,98}{\rm Ru}$ exceed the solar-abundance thresholds (blue arrows) only if mixing occurs at $\Delta r_{\rm mix}\lesssim 10\,$km, i.e., at smaller radii than for the $20\,M_\odot$ case, where $\Delta r_{\rm mix}\lesssim 20\,$km is sufficient. A similar conclusion applies to $^{92}{\rm Nb}$ production. In the absence of neutrino mixing, the predicted abundance lies well below the solar threshold (see the black diamond in the left panel of Fig.~\ref{fig:quantities_18.6}), whereas mixing at small radii (see, e.g., the red diamond corresponding to $\Delta r_{\rm mix}=1\,$km) brings agreement with observations.

Finally, the right panel of Fig.~\ref{fig:quantities_18.6} shows the production factors of the $p$-nuclides (purple dots) and $^{92}{\rm Nb}$ (green diamond with error bars), as defined in SM~D, for the unmixed case (upper panel) and for mixing at $\Delta r_{\rm mix}=1\,$km (lower panel). As in the benchmark model, $^{84}{\rm Sr}$ exhibits the largest production factor. Without neutrino oscillations, $\nu p$ yields beyond $^{84}{\rm Sr}$ are strongly suppressed, while with flavor equilibration at $\Delta r_{\rm mix}=1\,$km, $^{92,94}{\rm Mo}$ and $^{96,98}{\rm Ru}$ lie within a factor of $\sim 2$ below the co-production band (cyan region), indicating non-negligible production within uncertainties. The horizontal dashed orange line denotes the threshold required to reproduce solar abundances, corresponding to $f^{\rm ov}_{A,Z}=10$ (see Eq.~\eqref{eq:fov} and the discussion below). Mixing enhances the mass outflow rate $\dot{M}$ through increased heating, leading to a larger $M_{\rm out}$ and thus a slightly lower threshold in the $\Delta r_{\rm mix}=1\,$km case relative to the unmixed scenario~(see SM~D). Overall, production factors in the $18.6\,M_\odot$ model are smaller than in the benchmark $20\,M_\odot$ case (see Fig.~\ref{fig:Prod_factors}), implying that $p$-nuclides with $A>90$ are severely underproduced without mixing, while with mixing only species up to $^{98}{\rm Ru}$ exceed the solar threshold.

In summary, the $18.6\,M_\odot$ model has  reduced yields relative to the benchmark, yet effective neutrino mixing still supports efficient $\nu p$ process. Mo and Ru isotopes are produced at non-negligible levels---just below the co-production band yet above the solar threshold. In both models, mixing salvages the otherwise underproduced $^{92}{\rm Nb}$, putting it close to the co-production band (within $\sim1\sigma$) and satisfying observational constraints.